\documentclass[twocolumn]{aastex63}

\usepackage{url}
\usepackage{amsmath,amstext,amsfonts, amssymb}
\usepackage{parskip}
\usepackage{booktabs}
\usepackage{lipsum}
\usepackage{hyperref}
\shorttitle{gravity mode in accreting white dwarf}
\shortauthors{Kumar and Townsley}
\graphicspath{{./}{figures/}}

\begin{document}
\title{Gravity modes on rapidly rotating accreting white dwarfs and their variation after dwarf novae}
\correspondingauthor{Praphull Kumar}
\email{pkumar5@crimson.ua.edu}
\author[0000-0002-8791-3704]{Praphull Kumar}
\author[0000-0002-9538-5948]{Dean M. Townsley}
\affiliation{The University of Alabama,
 Tuscaloosa, Alabama, USA}
 
\begin{abstract}
Accreting white dwarfs in Cataclysmic variables (CVs) show short-period (tens of minutes) brightness variations that are consistent with non-radial oscillations similar to gravity (g) modes observed in isolated white dwarfs (WDs). GW Librae, a dwarf nova, was the first CV in which non-radial oscillations were observed and continues to be the best studied accreting WD displaying these pulsations. Unlike isolated WDs, accreting WDs rotate rapidly, with spin periods comparable to or shorter than typical low-order oscillation periods. Accreting WDs also have a different relationship between their interior temperature and surface temperature. The surface temperature of an accreting WD varies on a months to year timescale between dwarf novae accretion events, allowing study of how this temperature change effects g-mode behavior. Here we show results from adiabatic seismological calculations for accreting WDs, focusing on low-order ($\ell=1$) modes. We demonstrate how g-modes vary in response to temperature changes in the subsurface layers due to a dwarf nova accretion event. These calculations include rotation non-perturbatively, required by the high spin rate. We discuss the thermal history of these accreting WDs, and compare the seismological properties with and without rotation. Comparison of $g$-mode frequencies to observed objects may allow inference of features of the structure of the WD such as mass, surface abundance, accretion history, and more. The variation of mode frequencies during cooling after an outburst provides a novel method of identifying modes.

\end{abstract}

\section{Introduction} \label{sec:intro}
White dwarfs (WD) are stellar remnants that contain vital information about our galaxy and its various stellar populations, such as their age and star formation history \citep{Noh_1990, Garcia_2010, Rowell_2013, Kilic_2017, Fontaine_2021}. There are WDs in semi-detached binaries that are accreting material from their low-mass companion stars via Roche lobe overflow, cataclysmic variables \citep[CVs][]{Warner_1995}. Some of these systems have a mass transfer rate low enough to allow the detection of the WD photosphere \citep{Szkody_2002, Szkody_2007, Szkody_2012, Pala_2022}. A few of these systems, with directly observable WD photospheres, exhibit multi-periodic luminosity variations that are attributed to asteroseismic non-radial modes driven thermally by the surface convection zones \citep{Szkody_2016}. These pulsations resemble the $g$-modes observed in DA variables, also called ZZ Ceti type stars \citep[e.g.,][]{Brickhill_1991, Goldreich_1998, Saio_2013, Szkody_2021}.

Non-radial oscillations are sensitive to the star’s interior mechanical and thermal structure, hence asteroseismology offers the potential to probe aspects of the internal structure such as mass, radius, surface temperature, core temperature, extent of various compositional layers, and rotation behavior along with evolution history \citep{Kwaler_1985, Brassard_1992, Winget_2008, Althaus_2010, Romero_2012, Romero_2017, Corsico_2019}. Using asteroseismology on accreting white dwarfs (those in close binaries with mass transfer) is uniquely productive as it allows a probe of how the accretion of mass and angular momentum affects the WD and its subsequent evolution \citep{Sion_1995, Godon_2006}. Seismological variability of accreting systems can reveal the interior characteristics such as the mass of the accreted layer \citep{Townsley_Bildsten_2004, Townsley_Arras_Bildsten_2004}. 

The dwarf nova system GW Librae was the first CV to be observed with non-radial oscillations during the quiescence phase \citep{Warner_1998}, and has one of the shortest orbital periods, with a period of $P_{\rm orb}=77$ minutes \citep{Thorstensen_2002}. GW Lib showed mode periods at 648, 376, and 236 seconds. These modes had amplitudes and frequencies that remained consistent during the quiescent period for almost ten years \citep{Van_2004}. It also has a measured spin period of 209 seconds from spectroscopy \citep{Szkody_2012}. This is much shorter than the hours or days spin periods observed in isolated WDs and is very similar to the oscillation periods.

There is a large body of literature on the seismology of isolated WDs \citep[e.g.,][]{Althaus_2003, Corsico_2012, Althaus_and_corsico_2022}, however the current literature still lacks a complete seismological model on accreting WD systems. 
\cite{Townsley_Arras_Bildsten_2004} investigated the seismological structure of GW Lib without taking rotation into account. In preliminary work, \cite{Townsley_2016} was a non-refereed work submitted to be included in the proceedings from the conference ``The Physics of Accreting Compact Binaries" held in Kyoto, Japan in 2010. It is unclear if this prodeedings was ever published, and so the work was posted on arxiv.org in 2016. This makes the current work the first refereed work addressing g-modes in rapidly rotating accreting WDs. \cite{Townsley_2016} computed the $g$-mode behavior under rapid rotation but did not evaluate any time dependence that might arise from cooling after an accretion event, as we do here. Another major improvement in the current work over the models presented in \cite{Townsley_2016} is that the WD structures here are the result of evolution, performed with Modules for Experiments in Stellar Evolution \citep{Paxton_2011, Paxton_2013, Paxton_2015, Paxton_2018}, while those in \cite{Townsley_2016} are parameterized quasi-static models \citep{Townsley_Bildsten_2004}. \cite{Saio_2019} modeled and compared $r$-modes to a variety of objects, whereas here we consider $g$-modes in the limit of rapid rotation.

As the star spins up due to accretion,  it can have both $g$ and $r$ (Rossby)-modes, as well as modes that have different character in different parts of the star (mixed modes). The restoring force for $g$-modes is gravity, unlike that for $r$-modes, which is the coriolis force. For slowly rotating WD pulsators, the global $r$-modes are not important, notably not existing in the absence of rotation \citep{Kepler_1984}. Under particular approximations, it is possible to compute $r$- or $g$- modes but not mixed modes. Considering the observed pulsations as $r$-modes has met some success \citep{Mukadam_2013, Saio_2019}. However, more work is necessary because the utility of the necessary approximations in that work and this work is unclear.  We hope to pursue such an approximation-free treatment in future work, but for this work focus on $g$-modes in order to compliment the work of \citet{Saio_2019}.

Another aspect that we endeavor to include in our work here is an approach to a model of the accreting WD that accounts for the evolutionry history and ongoing accretion phenomenology.  While the actual evolutionary scenario that leads to the observed systems is still unclear \citep[e.g.][]{Goliasch_Nelson_2015,Schreiber_Zotorovic_Wijnen_2016, Shen_2022}, we have included the effect of accretion on the WD interior structure using constant accretion over long timescales.  We also compute, for the first time, the response of the seismological normal modes to the heating and cooling that takes place in response to the accretion coming in bursts.

DA or DB WDs pulsate on one instability strip depending on the type of enrichment in their atmospheres \citep[H or He,][]{Fontaine_2008}.  Due to their mixed surface composition, the accreting WDs may exhibit a single instability strip over a broader range of temperatures or a discrete set of instability strips.  The format will depend on the degree of helium enrichment, which is mostly dependent on the evolutionary status of the donor star \citep{Arras_Townsley_Bildsten_2006, Grootel_2015}. Unfortunately the amount of helium enrichment in any particular system is not easy to determine with precision.

Using spectroscopy, the observed width of the spectral lines can allow a determination of the surface rotation period \citep[e.g.,][]{Szkody_2012}. It is less clear how the surface rotation is coupled with the deep interior, and there is some hope that seismology may be able to address that question \citep{Hansen_1977}.  However, here we only consider uniformly rotating models.  Knowledge of WD masses in CVs, by contrast, has seen recent improvement with the advent of many precise distances from the \textit{Gaia} catalog \citep{Pala_2022}.  Since the number of observed modes can be rather few, broader seismological modeling of individual systems will benefit greatly from having masses constrained independently.

This paper is divided into the following sections: Section \ref{sec:wdhistory} provides the thermal history of the WD and describes constructing accreting WD models that are consistent with the observations. Section \ref{sec:seismologicalmethods} contains a discussion of the seismological methods for both rotating and non-rotating cases. Our results, the $g$-mode frequencies and eigenfunctions for the simulated stars and their variation during the cooling after an accretion event, are presented in section \ref{sec:results}. Section \ref{sec:discussion} discusses the results and the current limitations of this work, including planned future work. 

\section{White dwarf thermal structure and history} \label{sec:wdhistory}

In our study we used initial WD masses of 0.93~M$_{\odot}$ and 0.78 $M_{\odot}$. We provide a brief explanation of these models before moving on to the asteroseismology. Modeling was performed using Modules for Experiments in Stellar Evolution (MESA). Release version 10398 was used to perform the evolution of the models. The WDs were prepared in a way similar to that used in \cite{Timmes_2018}. We evolved models beginning at the pre-main sequence with initial masses 6.1$M_\odot$ and 3.95$M_{\odot}$ having solar composition (Z = 0.02) to a hot white dwarf (considering solid body rotation at ZAMS, $\Omega / \Omega_{crit}=1.9 \times 10^{-4}$), utilizing a 49-isotope reaction network. After winds reduced the hydrogen envelope mass below to $3\times 10^{-4} M_{\odot}$, we then stripped off all the remaining hydrogen from the surface. At this point the network is changed to $\texttt{CO\_burn}$. The WD is then cooled, incorporating element diffusion until the surface temperature reaches 15,000 K. Element diffusion is desired to smooth out the composition profile inside the star in order to make the profile of buoyancy frequency more smooth.

\begin{figure}
  \includegraphics[scale =1.0]{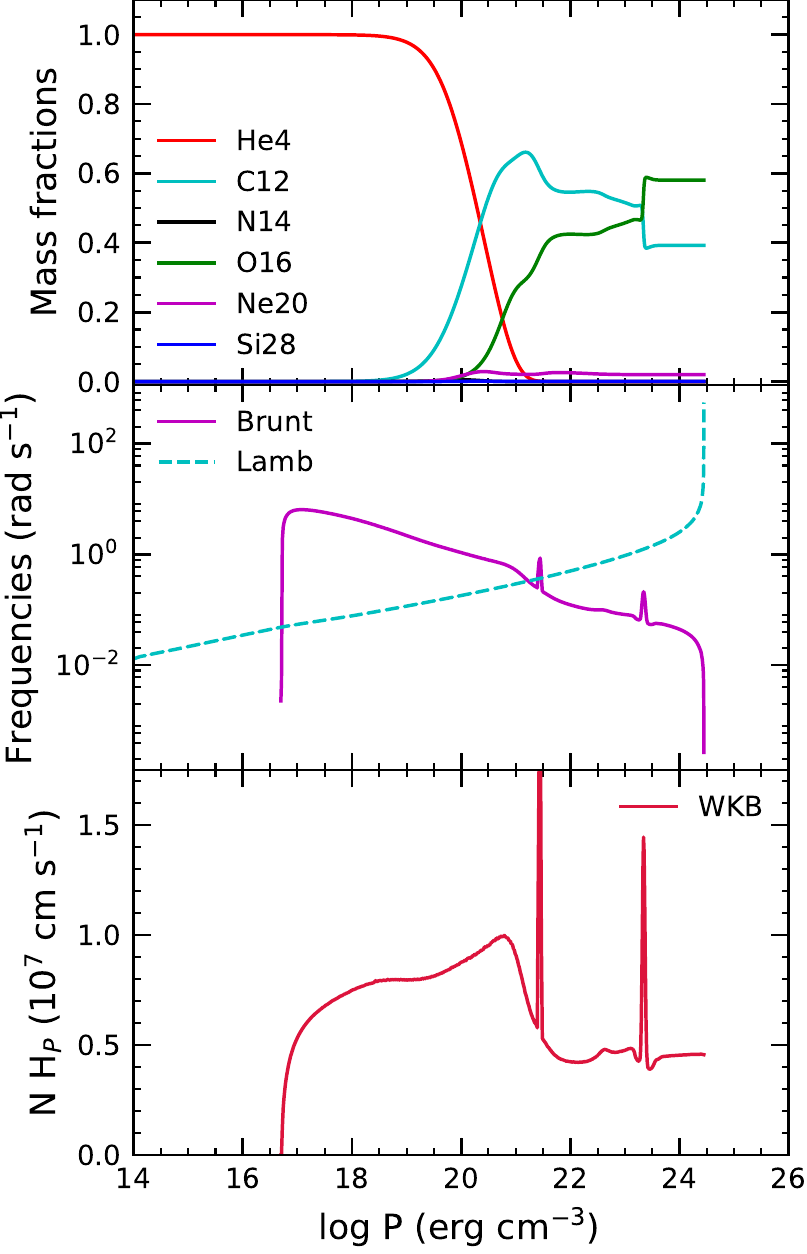}
   \caption{Composition (top panel), propagation diagram (middle panel), and WKB integrand (bottom panel) for 0.93~M$_{\odot}$ model at the end of cooling and before accretion.}
    \label{fig:wkb_integrand_brunt_abundance_after_cooling_0.93M} 
\end{figure}

The final WD models have masses M = 0.93~M$_{\odot}$ and 0.78~M$_\odot$, radii R = 0.0084 $R_{\odot}$ and 0.0101 $R_{\odot}$, an effective temperature $T_{\rm eff}$ of 15,000 K, and central temperatures $T_{c}$ = 1.01~$\times 10^{7}$ K and 1.28~$\times 10^{7}$ K respectively. Figure \ref{fig:wkb_integrand_brunt_abundance_after_cooling_0.93M} shows the profiles of the major abundances (top panel), Brunt-V{\"a}is{\"a}l{\"a} and Lamb frequencies (middle panel), and a WKB integrand (bottom panel) for the 0.93~M$_{\odot}$ model. The Brunt profile contains two peaks caused by composition gradients in the star. There is a broad bump between $\log P$ (erg cm$^{-3}$) of 20 and 21, due to the base of the He layer. The peak between $\log P$ of  $\sim$ 23 and 24 is due to the CO gradient at the edge of remnant of the convective core formed during core He burning. The sharp peak at $\log P\approx 21.5$ appears to be spurious and is a MESA artifact and the result of a problem with the equation of state and the composition in this region, this peak, while prominent here, is small compared to the feature that will be present once there is an accreted H layer, and it does not change during the accretion-cooling cycle that we are studying in this work. It will be addressed in future work. 

\begin{figure}
    \centering
    \includegraphics[scale =1.]{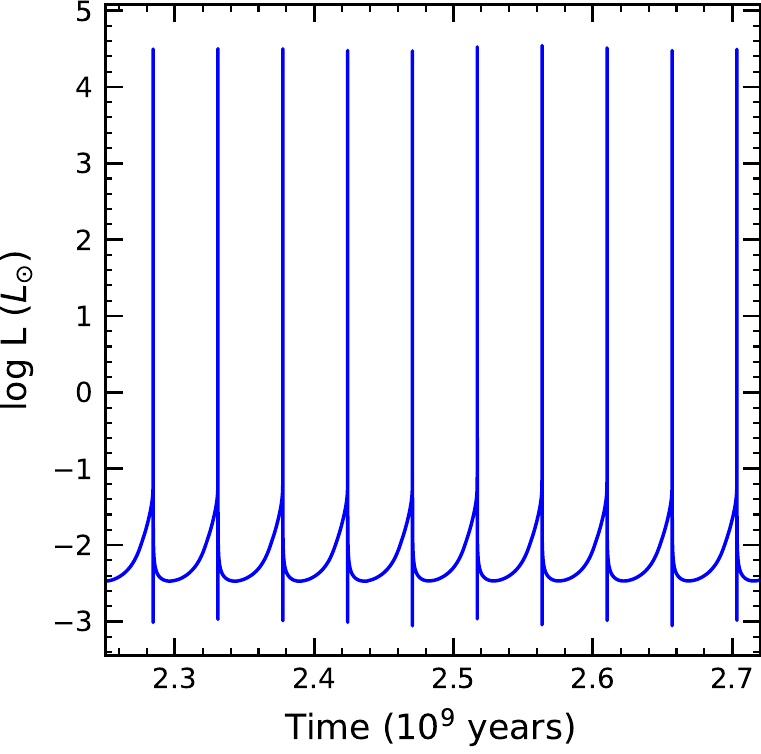}
    \caption{Luminosity history for 0.93~M$_{\odot}$ model during the classical novae cycle, shown here are the last 10 of 60 hydrogen flashes.}
    \label{fig:luminosity_profile_classical nova_0.93M}
\end{figure}

The cooled WDs (0.93~M$_{\odot}$ and 0.78~M$_\odot$ ) are subjected to a long-term accretion with  rate of 1.98~$\times$ $10^{-12}~ M_{\odot}~ yr^{-1}$. This accretion is applied for 2.76 Gyrs, during which the WD undergoes a total of 60 classical nova outbursts. Figure \ref{fig:luminosity_profile_classical nova_0.93M} demonstrates the luminosity history of the 0.93~M$_\odot$ model. The time here is taken as that which has elapsed since the onset of accretion, showing the last ten hydrogen flashes. The accretion rate is such that the typical median $T_{\rm eff}$ is near 14,400 K. The median is taken as the 50th percentile of the cumulative distribution of the time spent by the WD at each $T_{\rm eff}$ between consecutive hydrogen flashes. So that the WD spends about half of the time between nova outbursts
above and below this median $T_{\rm eff}$. The corresponding median luminosity is $\log L/L_\odot = -2.53$. The minimum luminosity $\log$ L~$\approx$~-2.6~L$_{\odot}$ and corresponding $T_{\rm eff}$ = 13,790 K. The accretion rate used in this work ($\dot M \sim 1.98\times10^{-12}$~M$_\odot$~yr$^{-1}$) is lower than what would usually correspond to this surface temperature for accreting WDs in CVs, which would typically be closer about an order of magnitude higher \citep{Townsley_and_Gansicke_2009, Pala_2022}. This comes about due to a few assumptions in the treatment of the long-term evolution of the accreting WD. All of these involve compromises made in order to make the evolution numerically more tractable. The core temperature used here ($T_c\approx 1.2\times 10^7$~K) is well above that expected for low accretion rates, which would be more like $5\times 10^6$~K \citep{Townsley_Bildsten_2004}.  Additionally, a very small mass fraction of $^{3}$He ( $\approx 3\times 10^{-5}$) is included in the accreted material, compared to the $\sim 10^{-3}$ expected for realistic donors, leading to an unrealistically late ignition of the classical nova, and more pre-ignition heating due to nuclear burning. Finally convective overshoot (dredge-up) is not included in the nova outburst phase. All of these work in the direction of increasing the contribution of other energy sources to the surface luminosity, thus decreasing the accretion rate needed to match the desired surface temperature. Ameliorating these deficiencies will form a critical part of future work, but do not affect the fundamental aspects of the mode structure created by the presence of rapid rotation, nor the variation caused by the cooling of the very outermost layers after a dwarf nova accretion event. Both of these WDs initially had helium layers before commencing accretion. The material being accreted is a solar mixture with photospheric mass fractions of hydrogen and helium of 0.7491 and 0.237, respectively \citep{Lodders_2003}.

\begin{figure}
    \centering
    \includegraphics[scale = 1.0]{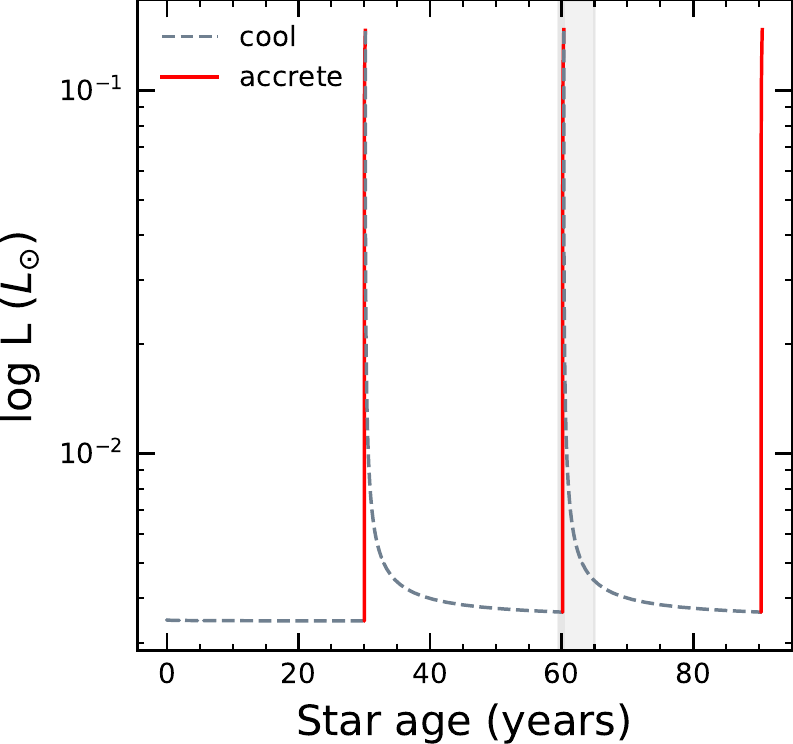}
    \caption{WD surface luminosity curve in response to dwarf nova accretion events for 0.93~M$_{\odot}$ model. The dashed grey lines indicate the quiescent time when the WD cools. The marked grey band is the region of interest for the seismological calculations.}
    \label{fig:luminosity_dwarf_nova_o.93M}
\end{figure}

\begin{figure}[ht!]
  \includegraphics[scale =1.]{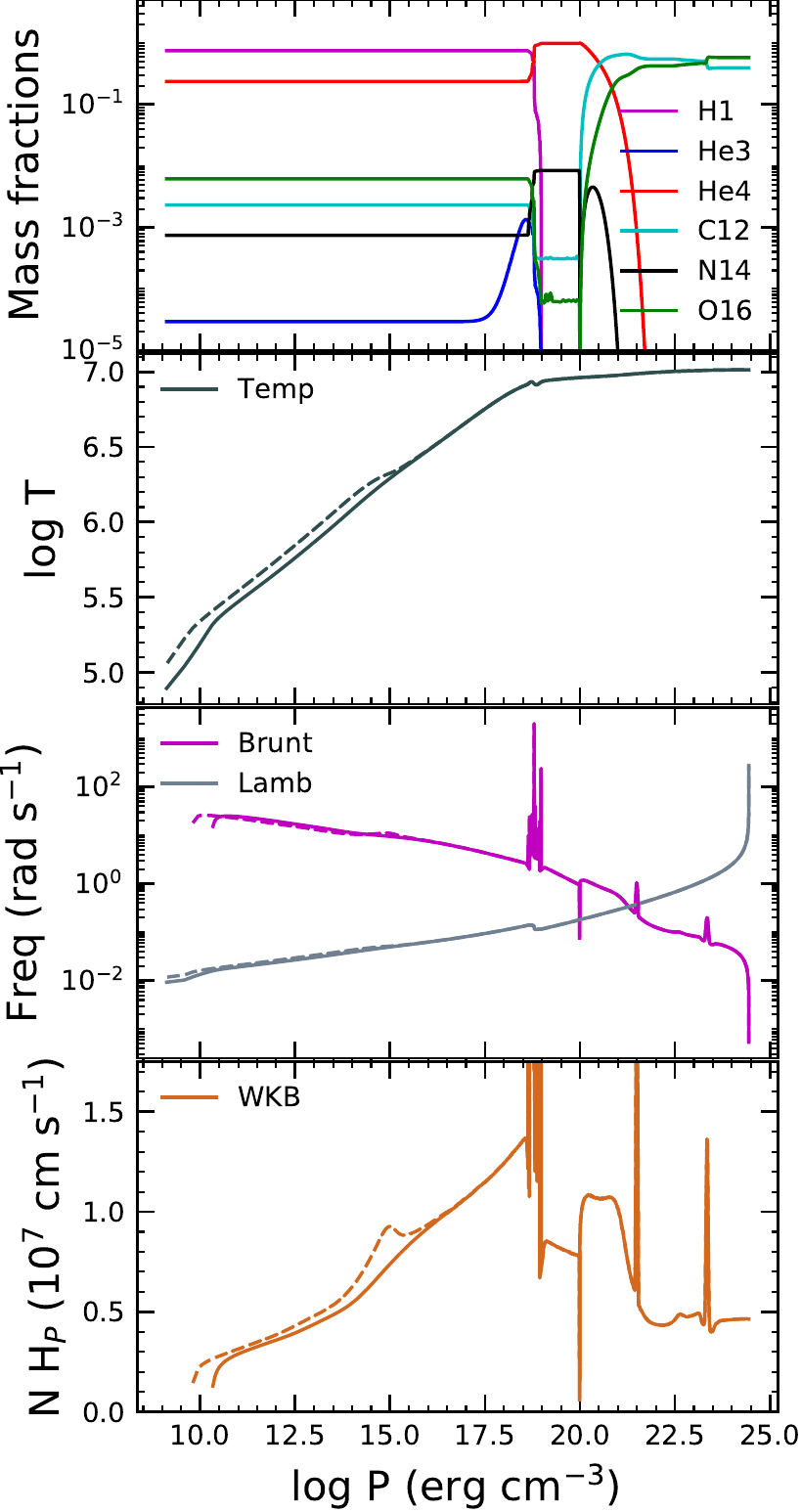}
  \caption{Demonstrated here are the abundances (top panel), Brunt and Lamb frequencies (middle panel), and temperature profiles (bottom panel), plotted against pressure, for 0.93~M$_\odot$ model. Solid and dashed lines represent the profiles plotted six months before and three months after the dwarf nova outburst respectively}.  
  \label{fig:profiles_before_after_dwarf_nova_0.93M}
\end{figure}

After completing approximately 40\% of the time interval between H flashes, we switch to modeling the dwarf nova accretion cycle in detail rather than using continuous accretion at a time-averaged value. We apply a sequence of dwarf nova accretion events with recurrence time of 30 years, allowing the accretion to be active for two months at an accretion rate of 1.2~$\times$ $10^{-8} ~M_{\odot} ~yr^{-1}$. Figure \ref{fig:luminosity_dwarf_nova_o.93M} shows the light curve. To demonstrate the interior structure relevant to seismology, Figure \ref{fig:profiles_before_after_dwarf_nova_0.93M} shows the variation of the Brunt-v\"ais\"al\"a along with the Lamb frequencies (third panel from top). The solid and dashed lines in Figure \ref{fig:profiles_before_after_dwarf_nova_0.93M} represent the WD structure six months before and three after the outburst. The top panel shows the abundance profile six months before the outburst, and the bottom panel is the WKB integrand. The contrast shown indicates that the frequency change will be small but measurable. The negative bump in the Brunt frequency at $\log P\sim 20$ in the middle panel of Figure \ref{fig:wkb_integrand_brunt_abundance_after_cooling_0.93M}is due to a local inversion of the mean molecular weight at the boundary between the accreted material and the starting WD model. As can be seen from the abundances shown in the top panel, this point corresponds to the location where the heavier N/O ashes from CNO hydrogen burning during flashes is layered on top of the lighter C/O. This boundary is artificially sharp due to our not including element diffusion during accretion, which we hope to improve upon in future work.

Changes in the WKB integrand translate directly to changes in the frequency of a mode of fixed order.
This is helpful for interpreting our results, which have to do with how the heating and cooling of a modest outer region of the star will change mode frequencies.
The WKB integral is a phase accumulation for modes in the high-order limit, and the integrand that appears in this integral is a better indication of the impact of the local structure on mode propagation than the Brunt-V\"ais\"al\"a frequency alone.
Within the WKB approximation, a plane wave form is assumed with a radius-dependent wavenumber $k_r$.
As a result, the phase $\phi$ accrued by the wave in a given region is $\phi$ = $\int$ dr $k_{r}$ =$[\ell(\ell+1)]^{1/2}\omega^{-1}\int N\;dr/r$ where $N$ is the Brunt-V\"ais\"al\"a frequency, $\ell$ is the angular order of the mode, and $\omega$ is the frequency of the mode.
So the WKB integrand is proportional to the number of nodes per decade in pressure inside the star.
For a mode of fixed radial order, this can be solved for $\omega$ so that, approximately, $\omega \propto (1/R) \int NH_{P}\;d\ln P$, where $H_P$ is the pressure scale height.
Thus, the integrand reflects how changes in local properties influence the frequency of a mode.

\section{Seismological Methods} \label{sec:seismologicalmethods}
To determine the asteroseismic properties of our white dwarfs, we use version 5.2 of the open source stellar oscillation code Gyre \citep{Townsend_2013, Townsend_2018}. In this paper, we only consider the adiabatic evaluation of non-radial gravity modes of the star, although Gyre can also compute non-adiabatic contributions.

While the documentation for Gyre gives a full description of the equation it solves and other work is more appropriate for a broad overview of seismology \citep{Unno_1989}, we will briefly summarize here the essential basis of the seismological calculation. An important distinction between WDs and accreting WDs is that accreting WDs are likely to be rapidly rotating. This has broad and critical impacts on how seismology is performed. Beyond a summary of the calculation being performed, we will orient our discussion around distinguishing the treatment of the non-rotating situation compared to the rapidly rotating situation. The seismology of many isolated WDs can be studied by treating the modest rotation as a perturbation. This is not possible for the rotation rates expected for accreting WDs, though there are well-known methods for performing seismological analysis with rapid rotation (i.e., outside the perturbative limit).
\subsection{Non-rotating WD seismology} \label{ssec: nonrotating wd seismology method}
Assuming that the stellar structure is nearly static and the materials inside are inviscid. The general equations that needed to solve  are the conservation of mass, conservation of momentum, Poisson's equation, and energy equation. The governing equations are
\begin{eqnarray}
\label{eq:1}
\frac{\partial\rho}{\partial t} + \vec{\nabla}\cdot\left(\rho\vec{v}\right) &=& 0\ , \\
\label{eq:2}
\rho\left(\frac{\partial}{\partial t} + \vec{v}\cdot\vec{\nabla}\right)\vec{v}
&=& -\vec{\nabla} P + \rho\vec{g}\ , \\
\label{eq:3}
\nabla^{2}\Phi &=& 4\pi G \rho\ , \\
\label{eq:4}
\frac{\partial P}{\partial t} + \vec{v}\cdot \nabla P &=& c_{s}^{2}\left(\frac{\partial \rho}{\partial t} +\vec{v}\cdot\nabla\rho\right)\ .
\end{eqnarray}
Where $\rho$, $v$, and $P$ are the density, velocity, and pressure of the stellar material, respectively. $\vec{g}$ and $\Phi$ are the gravitational acceleration and gravitational potential. These two are related as $\vec{g} = -\nabla\Phi$. $c_s$ is the adiabatic speed of sound. Now consider a small perturbation that displaces the above quantities from hydrostatic equilibrium,

\begin{eqnarray}
\label{eq:5}
\rho &\rightarrow & \rho_{0} +\rho ^{\prime}\ , \\
\label{eq:6}
\vec{v} &\rightarrow & \vec{v}_0 +\vec{v}^{\prime}=\frac{d\vec{\xi}}{dt}\ , \\
\label{eq:7}
P  &\rightarrow & P_{0} + P^{\prime}\ .
\end{eqnarray}
Subscript 0 denotes the quantities at equilibrium state, and primed variables indicate the Eulerian perturbations about the equilibrium. $\vec{\xi}=\vec{r}-\vec{r}_0$ is the Lagrangian displacement in space. At equilibrium, the system is in hydrostatic equilibrium therefore $-\vec{\nabla}P_{0} +\rho_{0}\vec{g} = 0$ and velocity field is $\vec{v}_0=0$ .

We will consider linear, adiabatic oscillations, for which the Lagrangian perturbation of the entropy is zero. The pulsation equations for a non-rotating and, non-magnetic star will be in terms of the Eulerian perturbations.
Assuming spherical symmetry, the displacement vector ($\vec{\xi}$) can be separate out into the radial and horizontal components, $\vec{\xi} = \xi_r\hat{r}+\vec{\xi}_h$ with $\vec{\xi}_h =\xi_{\theta}\hat{\theta}+\xi_{\phi}\hat{\phi}$. Considering only oscillatory the form of the solutions the radial displacement can be taken proportional to Y$_{lm}(\theta,\phi)e^{-i\omega t}$. Where Y$_{l}^{m}$ is Laplace's spherical harmonic for degree $l$ and $m$. We ignore the perturbation in gravitational potential. This is often called the ``Cowling approximation" \citep{Cowling_1941} and reduces the complexity of the problem. This approximation has only a small effect on the mode frequency calculations for the non-rotating star. If the Cowling approximation is not used, the mode frequency would change by approximately 0.02\% as evaluated using Gyre with the approximation disables. We will omit subscript 0 for the equilibrium quantities. The final linearized equations are \citep{Aerts_2010, Unno_1989} 
\begin{eqnarray}
\label{eq:8}
\frac{dP^{\prime}}{dr} &=& \rho\left(\omega^{2}-N^{2}\right)\xi_{r} + \frac{1}{\Gamma_{1}P}\frac{dP}{dr}P^{\prime}\ , \\
\label{eq:9}
\frac{d\xi_{r}}{dr} &=& -\left(\frac{2}{r}+\frac{1}{\Gamma_{1}P}\frac{dP}{dr}\right)\xi_{r} + \frac{1}{\rho c_{s}^{2}}\left(\frac{S_{l}^{2}}{\omega^{2}}-1\right)P^{\prime}\ ,
\end{eqnarray}
where $c_{s}^{2} = \Gamma_{1} P/\rho$ is the squared adiabatic sound speed and $\Gamma_{1}$ = $\left(\frac{dlnp}{dln\rho}\right)_{ad}$  is the adiabatic exponent, $N^{2}=-g\left(d\ln \rho/dr+g/c_{s}^{2}\right)$ is the squared Brunt-v\"ais\"al\"a frequency or buoyancy frequency, $S_{l}^{2} = l(l+1)c_{s}^{2}/r^{2} = k_{h}^{2}c_{s}^{2}$ is the Lamb frequency and $k_{h}$ is the horizontal wave number. The density perturbation can be expressed entirely in terms of $\xi_r$ and $P'$
\begin{eqnarray} \label{eq:10}
\rho^{\prime} = \rho \left(\frac{P^{\prime}}{\Gamma_1 P} + \xi_{r} \frac{N^2}{ g}\right).
\end{eqnarray}

In order to ease the solution, Gyre re-frames the variables slightly by using the fractional radius, $x=r/R$, as the independent variable and scaling the functions $\xi_r$ and $P'$ to create the dimensionless functions (see the Gyre documentation)
\begin{eqnarray}
\label{eq:11}
y_1 &=& x^{2-l}\frac{\xi_{r}}{r}\ , \\
\label{eq:12}
y_2 &=& x^{2-l}\frac{P^{\prime}}{\rho g r}\ . 
\end{eqnarray}

\subsection{Rotating WD seismology} \label{ssec: rotating wd seismolgy method}
For an isolated WD, the ratio of centrifugal to gravitational acceleration is very small,  typically on the order of $\sim 10^{-2}$ or smaller, and the spin frequency is a similarly small fraction of the typical observed mode frequency, therefore the rotational force is considered a weak perturbation. However, in rapidly rotating CV WDs spun up by accretion, the spin frequency of the star becomes similar to or larger than the mode frequency. Thus, rotation can not be treated in a perturbative fashion, as the spin will break the spherical symmetry of the problem. The Coriolis force must be included in the momentum equation that is solved for normal modes. The method to treat these stars comes from  geophysics and is called the ``Traditional approximation (TAR)" \citep{Chapman_and_lindzen_1970, Bildsten_Ushomirsky_Cutler_1996}. The approximation maintains the separability of the normal mode problem into radial and angular parts. Its applicability here does have some limitations, discussed below. 
Consider a WD which is uniformly rotating and non-magnetic. We work in the corotating frame (in which fluid is at rest) and assume a periodic time-dependence for the pulsation, proportional to $e^{i\omega t}$. The linearized momentum equation for a rotating star is given by:
\begin{eqnarray} \label{eq:13}
i\omega\frac{d\xi}{dt} &=& -\frac{\nabla P}{\rho} - g \hat{r} - 2i\omega\vec{\Omega} \times \vec{\xi} 
\end{eqnarray}
Here the last term corresponds to the Coriolis force and $\Omega$  is the rotation frequency of the star and $\omega$ is the mode frequency in the rotating frame.

The perturbed momentum equation becomes \citep{Bildsten_Ushomirsky_Cutler_1996,Lee_and_Saio_1997}:
\begin{eqnarray}\label{eqn}
-\rho \omega^2 \xi_{r} -2 i\rho\omega\Omega \sin \theta \xi_{\phi} + g \rho^{\prime} &=& -\frac{\partial P^{\prime}}{\partial r} \label{eq:14} \ , \\
-\rho \omega^2\xi_{\theta} - 2 i \rho\omega\Omega \cos \theta\xi_{\phi}&=& -\frac{1}{r}\frac{\partial P^{\prime}}{\partial\theta} \label{eq:15} \ , \\ 
\nonumber
-\rho \omega^2 \xi_{\phi} + 2 i\rho\omega\Omega \cos \theta\xi_{\theta} + &&\\ 2 i \rho \omega \Omega \sin \theta \xi_{r} &=& 
-\frac{1}{r \sin \theta}\frac{\partial P^{\prime}} {\partial \phi }\ . \label{eq:16}
\end{eqnarray}

Using eqn. (\ref{eq:10}), we can reduce eqn. (\ref{eq:14}) to the form, 
\begin{eqnarray}\label{main equation}
\rho\left(\omega^2 - N^2\right)\xi_{r} &=& \frac{\partial P^{\prime}}{\partial r} + \frac{\rho ^{\prime}}{h\Gamma_1} - 2i\omega\Omega\rho\sin \theta \xi_{\phi }\ .
\end{eqnarray}

Where $h = P/\rho g$ is the local pressure scale height. 
For the case of low-frequency oscillations and assuming that the waves are almost incompressible \citep[see][]{Unno_1989}), $\mathbf{k}\cdot \mathbf{\xi} \approx 0$, we obtain
\begin{equation} \label{eq:18}
\frac{\xi_{r}}{\xi_{h}}\approx -\frac{k_h}{k_r}
\end{equation}
If $|N^2|$ is much larger than $\omega^2$ and $\Omega ^2$, the dispersion relation reduces to $k_{h}^{2}/k^{2}\sim 0$, leading to us to consider the horizontal component of the wave number negligible as compared to radial wave number, indicating $\xi_{h}$ $\gg$ $\xi_{r}$. Therefore, the term containing $\xi_r$ in equation (\ref{eq:16}) can be neglected for low $\ell$, low frequency oscillations. 
Using similar arguments, $\xi_{\phi}/\xi_{r} \sim r/h$. Thus the Coriolis term in equation (\ref{main equation}) can be neglected when $N^2 \gg r\Omega\omega /h$. This approximation is known as \textit{"Traditional approximation"}.
 
This approximation greatly simplifies the solution of the above perturbative equations, as it separates out the radial and angular part completely. It simplifies equation (\ref{main equation}) to be same as equation (\ref{eq:8}), which appeared in the non-rotating problem. We can see this by combining equations (\ref{eq:15}) and (\ref{eq:16}), and writing $\xi_{r}$ and $\xi_{\theta}$ in terms of $\delta p$ and $\xi_{r}$. The solutions within the TAR can be given in terms of $\xi_{r}$ and $\xi_{h}$, where $\xi_{h}$ is the horizontal part of the displacements, which can be decomposed to $\xi_{\theta}$ and $\xi_{\phi}$ respectively (see Gyre documentation for the general forms). 

The final equations solved for the radial dependence of the eigenfunctions
are the same as for the non-rotating case, i.e. Equations (\ref{eq:8}) and (\ref{eq:9}), except that $\ell(\ell+1)$ is replaced by the eigenvalue, $\lambda$, of Laplace's
tidal equation, given as: \citep{Bildsten_Ushomirsky_Cutler_1996}:
\begin{eqnarray}\label{eqn:19}
\nonumber
   \left[ \frac{\partial}{\partial\mu}\left(\frac{1-\mu^2}{1-q^2}\frac{\partial}{\partial\mu}\right) - {} \right. &&\\ \left. \frac{m^2}{(1-\mu^2)(1-q^2\mu^2)} - \frac{qm(1+q^2\mu^2)}{(1-q^2\mu^2)^2}\right]\Theta &=& -\lambda \Theta
\end{eqnarray}
with $\mu=cos \theta$. $\Theta$ is the Hough function, which is the solution to this equation, and $\Theta(\theta) e^{im\phi}$ replaces $Y_\ell^m(\theta, \phi)$ in the full solutions. Thus, reassembling the solutions after separation of variables makes, for example, the radial displacement $\xi_r(r)\Theta(\theta)e^{im\phi}e^{i\omega t}$.  A similar construction applies to $P'$. One can visualize $\lambda$ as a transverse wave number. The solutions of Laplace's tidal equation have dependence on the spin parameter $q=2\Omega/\omega$ and on $\ell$ and $\mathrm{m}$. The $\phi$ and time dependence of all eigenfunctions continues to be $e^{im\phi}\times e^{i\omega t}$.

Now, for the boundary conditions, we consider regularity at the center, i.e., r = 0 (where the solutions are continuously differentiable and that prevents any singularity in the problem). This can be seen by the following arguments: near the origin, the equations (\ref{eq:8}) and (\ref{eq:9}) are approximated as $d\xi_r/dr = (\ell(\ell+1)/\rho \omega^2 r^2) P'$ and $dP'/dr = \rho \omega^2 \xi_r$; so that $d^2P'/dr^2 \approx \rho \omega^2 d\xi_r/dr = (\ell(\ell+1)/r^2)  P'$. Then doing a power series expansion, the solutions for $P'$ and $\xi_r$ can be obtained as: $ P' = A r^{(\ell+1)}$ and then $\xi_r = (1/\rho \omega^2) A (\ell+1)r^{\ell} = A((\ell+1)/\rho \omega^2) r^{\ell} $. This is consistent with the boundary equations in the gyre documentation. Note that since there is only a relation between $\xi_r$ and $P'$ specified, this allows $\mathrm{\omega}$ to be solved as an eigenvalue and leaves the normalization (amplitude) as a free parameter.

At the star's surface, we assume that the Lagrangian pressure perturbation vanishes, so that the Eulerian pressure perturbation is given as $P^{\prime} = \rho g\xi_r$ ($\delta P=0$). Frequencies in non-rotating frame are given by $\omega_{i} = \omega + m \Omega$, where $\mathrm{\omega}$ is the oscillation frequency in the corotating frame, $\mathrm{\Omega}$ is the angular frequency of the rotation of the star, and $\mathrm{m}$ is the eigenvalue in the azimuthal direction of Laplace's tidal equation. 

\subsection{Solution methods}
The linearized pulsation equations in the adiabatic case are basically two-point boundary value problems (BVP). Gyre uses the Magnus Multiple Shooting (MMS) scheme to solve the linearized pulsation equations. The MMS scheme considers boundary value problems as a set of initial value problems (IVPs), in which initial values are adjusted to match a second boundary condition. Normalization is used to avoid the singularities in the discriminant function. That leads to dividing by one of the dependent variables calculated at the boundary \citep{Unno_1989}. The MMS method used in Gyre avoids the exponential dichotomy problem, in which the preferred integration or differencing direction for some terms is opposite that of others. That has particular advantages for non-adiabatic calculations, although here we only perform adiabatic calculations. In work not utilizing Gyre, a common method is the relaxation scheme, which replaces the derivatives with finite-difference approximations on a specified grid. This approach \citep{Castor_1971} has been used for numerous oscillation calculations. 

The MMS scheme solves the overall BVP on the domain by dividing into subintervals, a grid of $N$ points $x^{a} = x^{1} < x^{2} < ...< x^{N} = x^{b}$. It then solves the BVP in each subinterval with $N-1$ matching conditions using the boundary conditions described in \cite{Townsend_2013}[cf. eq. 2]. The IVP is then solved using Magnus integrators. At the lowest order, this approach is based on the approximation of the Jacobian matrix of the adiabatic pulsation equations as a piecewise constant in each subinterval \citep{Gabriel_and_Noels_1976}. To extend to higher order, Gyre uses the Gauss-Legendre quadrature to solve the Magnus matrix integral, with various orders of accuracy, e.g., GL2, GL4, GL6. We use the fourth order (GL4) of accuracy. Although Gyre has its own grid construction, it can also be cloned from the stellar input model with the option of oversampling the subintervals. We use an \texttt{INVERSE} grid type for g-mode frequency search ensuring at least 10 points per wavelength in the propagation region, 5 points per scale length in the evanescent region, and 5 points between the center and the inner turning points. This also allows Gyre to have enough room to search for most of the $g$-modes frequencies in the propagation region.  

\section{Results} \label{sec:results}
Here we show the seismological calculations of the WDs using Gyre for both non-rotating and rotating cases of two different mass models, 0.93~M$_{\odot}$ \& 0.78~M$_\odot$.
Results for the non-rotating calculation are shown first. These can be compared to earlier work on accreting WDs \citep{Townsley_Arras_Bildsten_2004} as well as to the broader work on isolated white dwarfs. This is followed with calculations of seismology in the rapid rotation limit, using the observed spin frequency of GW Lib as a typical value. This is specifically applicable to accreting WDs, as the effect of spin on the modes cannot be treated perturbatively as isolated WDs can. Finally, we present results for how the heating of the outer layers introduces small but potentially observable changes to the normal g-mode frequencies that also change characteristically in time.

\subsection{Non-rotating seismology}
\label{ssec:non-rotating results}
\begin{figure}[hb!]
   \includegraphics[scale = 1.05]{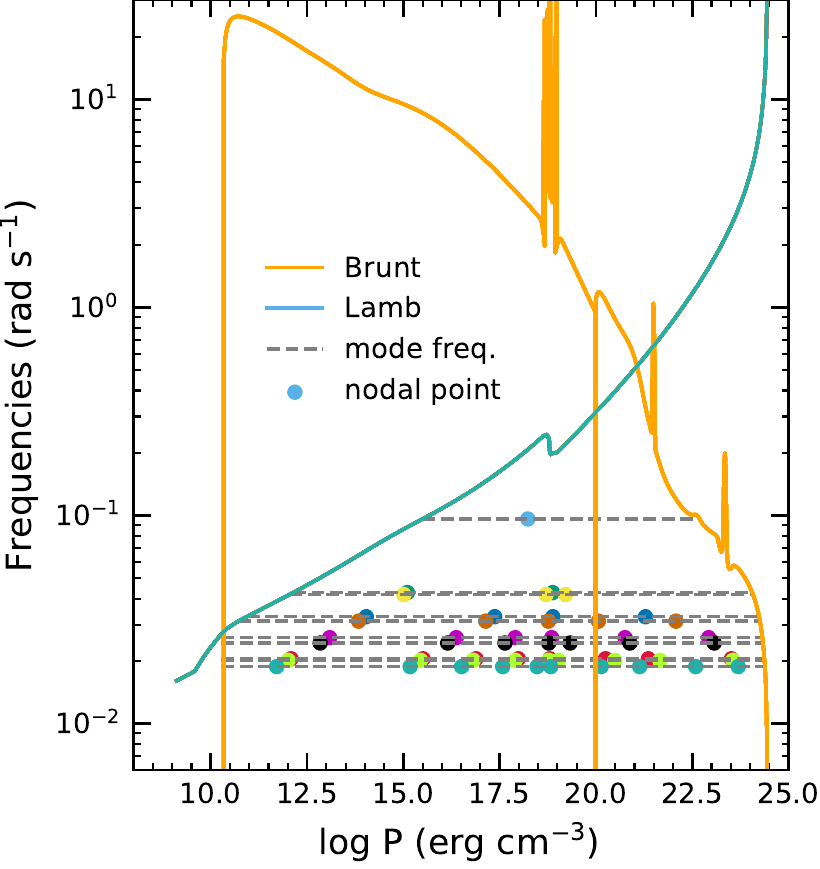}
   \caption{Propagation diagram for the 0.93~M$_{\odot}$ WD model for the non-rotating case six months before the accretion event. The solid orange line is the Brunt-V{\"a}is{\"a}l{\"a} frequency, while the solid green shows the $\ell=1$ Lamb frequency. The frequencies of the ten lowest order $g$-mode are indicated by dashed horizontal lines. Pressure location of the zero crossings of the radial displacement $\xi_r$ are indicated with solid dots of varying color. The surface is on the left in this coordinate.\label{fig: propagation_diagram_6mb_norot_0.93M} }
\end{figure}

In this section, we show the results of the low-order ($\ell=1$) non-radial normal modes in a non-rotating white dwarf. Figure \ref{fig: propagation_diagram_6mb_norot_0.93M} shows the propagation diagram of a 0.93~M$_\odot$ model six months before the dwarf nova accretion event and 30 years after the previous accretion event. The solid orange line indicates the Brunt-V\"ais\"al\"a frequency, while the solid green line is the Lamb frequency. These two characteristic frequencies determine the propagation regions of two different modes: modes propagating with larger frequencies than both are identified as pressure modes, whereas modes propagating with smaller frequencies than both are considered as gravity modes. Modes observed in white dwarfs are dominated by $g$-modes \citep[][and references therein]{Winget_2008, Corsico_2020}. Dashed grey lines indicate the frequencies of the first ten non-radial gravity modes from top ($n_{g} = 1$) to bottom ($n_{g} = 10$). The horizontal axis represents the logarithm of the pressure coordinate and can be transformed to fractional depth into the star, $1-r/R$.
 
Whereas the propagation diagram indicates the regions of the star in which a mode of a particular frequency can propagate, inspecting the mode eigenfunctions directly provides more insight into how propagation and mode amplitude differ in various parts of the star. The contrast between the behavior in the core and envelope is particularly important. The deep well in fig \ref{fig: propagation_diagram_6mb_norot_0.93M} at $\log P \sim 20$ is due to an inversion of mean molecular weight, when the heavier N/O ashes from the hydrogen flashes is layered on top of the lighter C/O.

\begin{figure}[ht!]
   \includegraphics[scale = 1.08]{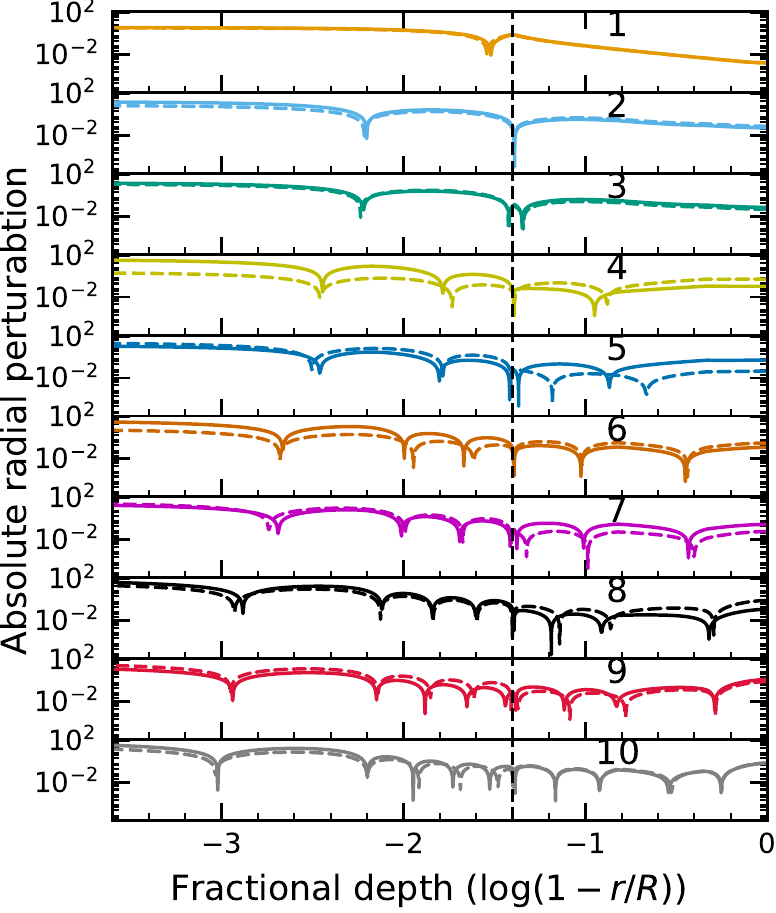}
   \caption{Eigenfunctions for absolute radial displacement, $|\xi_r|$, for the first ten modes of for $\ell=1$. The solid line  indicates the eigenfunctions three months after the accretion outburst and the dashed line shows the eigenfunctions six months before the accretion outburst (approximately 30 years after the previous one) of a non-rotating 0.93~M$_{\odot}$ white dwarf model. Vertical dashed line indicates the boundary of core-envelope. The surface is on the left in this coordinate.\label{fig: eigenfunction_6months_before_and_three_months_after_norotation_0.93M} }
\end{figure}

Figure \ref{fig: eigenfunction_6months_before_and_three_months_after_norotation_0.93M} shows the radial displacement eigenfunctions, $|\xi_r|$, of the first ten $g$-mode of 0.93~M$_{\odot}$ model. The horizontal axis represents the logarithm of the fractional depth into the star, $1-r/R$. The right side of Figure \ref{fig: eigenfunction_6months_before_and_three_months_after_norotation_0.93M} is at the core of the star. The solid line and dashed line indicate the radial perturbation three months after and six months before the accretion outburst, respectively, for each order. The eigenfunctions computed by Gyre are normalized such that each mode has inertia equal to $M R^2$ \citep{Aerts_2010, Townsend_2013}. It can be seen that while most of the node locations change only modestly due to the surface heating resulting from the accretion, the nodes in the core for the $n_g=5$ mode shift quite significantly. This is expected, as a modest change at the boundary can lead to an overall change in the resonant normal mode if the mode structure works out just right. In this case, one of the nodes moves from the boundary layer between the light element's (H-rich) shell and the (H-poor) core to the interior of the core.

An apparent ``pairing" of consecutive modes (a pronounced alternating small-large frequency spacing) is evident in Figure \ref{fig: propagation_diagram_6mb_norot_0.93M} and in results presented below, including cases with rotation, though it is less pronounced for the 0.78~M$_\odot$ model than the 0.93~M$_\odot$ model. In the absence of internal structured layers in the WD, $g$-modes are expected to have an approximately uniform spacing in period and therefore higher order modes appear closer together in frequency. Departures from this regular period spacing are introduced by the temperature and composition structure of the WD. As the mode order is increased, new radial nodes must appear either in the core or the envelope. This typically leads to alternation between smaller and larger spacings between consecutive modes \citep{Bradley_1993}, though not always the strict odd-even alternation seen here. This effect appears exaggerated here. We believe this is due to two causes: First, the WKB integrated phase, the integral of the function shown in the lower panel of Figure 
\ref{fig:profiles_before_after_dwarf_nova_0.93M}, in the core and the envelope are comparable. This means that every other mode will add one node in the core and one in the envelope. Observing the eigenfunctions of the even-order modes (2, 4, 6, 8, 10) in Figure \ref{fig: eigenfunction_6months_before_and_three_months_after_norotation_0.93M} confirms this expectation. The level of regularity is unusual compared to cooling WDs, which don't usually follow a strict core-envelope alternation. The second cause then becomes important: usually this alternation would mean that the additional node for the intervening modes (3, 5, 7, 9) would appear in either the core or the envelope. However, due to our not including element diffusion during accretion, the boundary between the core and envelope is excessively sharp, giving a very strong peak in the Brunt frequency (at $\log P \sim 18.5$). From the eigenfunctions shown in Figure \ref{fig: eigenfunction_6months_before_and_three_months_after_norotation_0.93M}, this appears to have caused the added node to appear at the boundary rather than in the core or the envelope. Such a placement appears to exaggerate the non-uniformity of the mode spacing, making the smaller period spacing much less than the larger, and mode frequencies to appear in fairly close pairs.

\subsection{Rotating seismology}
\label{ssec:rotating results}
When the spin frequency and mode frequency become comparable, the ``splitting'' normally associated with the influence of rotation on modes becomes comparable to the mode frequency. This is expected to be common for $g$-modes in CV WDs that have undergone a long history of accretion, thus having the opportunity to gain a large amount of angular momentum. Here we will show how rotation has significant impacts on the mode structure and eigenfunctions. For examples where we must choose a specific spin, we consider that the spin period of the star is similar to the observed GW Lib spin period, 209 seconds \citep{Szkody_2012}, which is much shorter than the hours or days observed in isolated WDs.

\begin{figure*}[ht!]
\includegraphics[scale = 0.98]{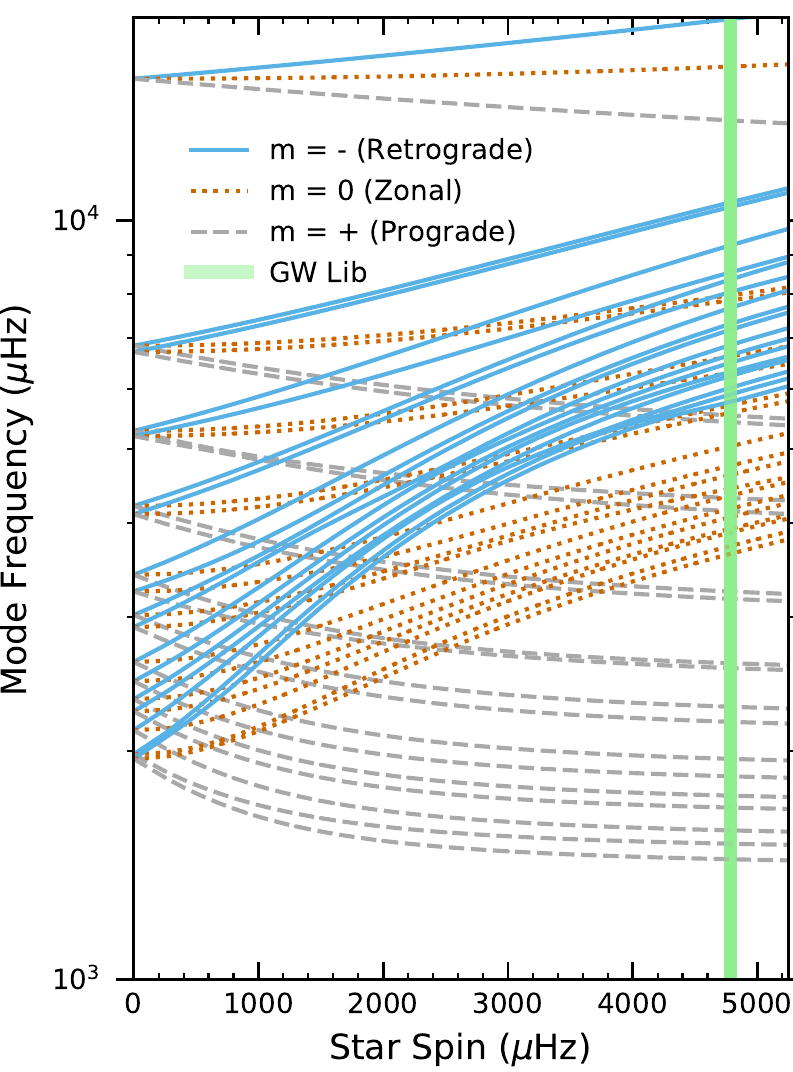}
\includegraphics[scale = 0.98]{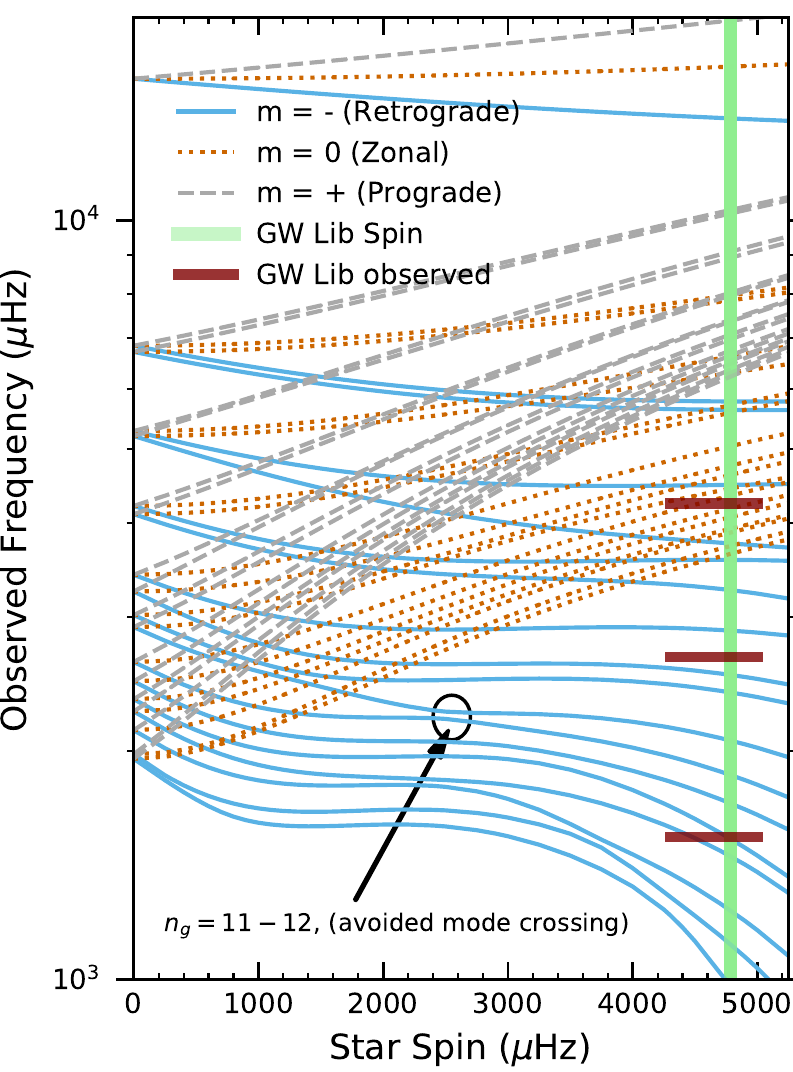}
\caption{First 18 $g$-mode (topmost is the lowest radial order mode) frequencies as a function of star’s spin frequency calculated for $\ell$ = 1 for the 0.93~M$_{\odot}$ model, three months after the dwarf nova accretion event. The left panel shows the frequency in the (rotating) star’s frame and the right panel shows the frequency in the non-rotating frame, that is seen by a fixed observer, $\omega_i = \omega +m\Omega$. Solid blue lines are the retrograde modes ($m=-1$), dashed grey lines are the prograde modes ($m=1$), and dotted crimson lines are the zonal modes ($m=0$). Observed GW Lib spin frequency is indicated by the solid green line. The frequencies observed in GW Lib are indicated by the solid maroon horizontal lines on the right panel. From top to bottom: 4237 $\mu$Hz, 2660 $\mu$Hz, and 1543 $\mu$Hz.}
\label{fig:modes vs spin frequency for 0.93M after three months}
\end{figure*}

\begin{figure}[hb!]
\includegraphics[scale = 0.98]{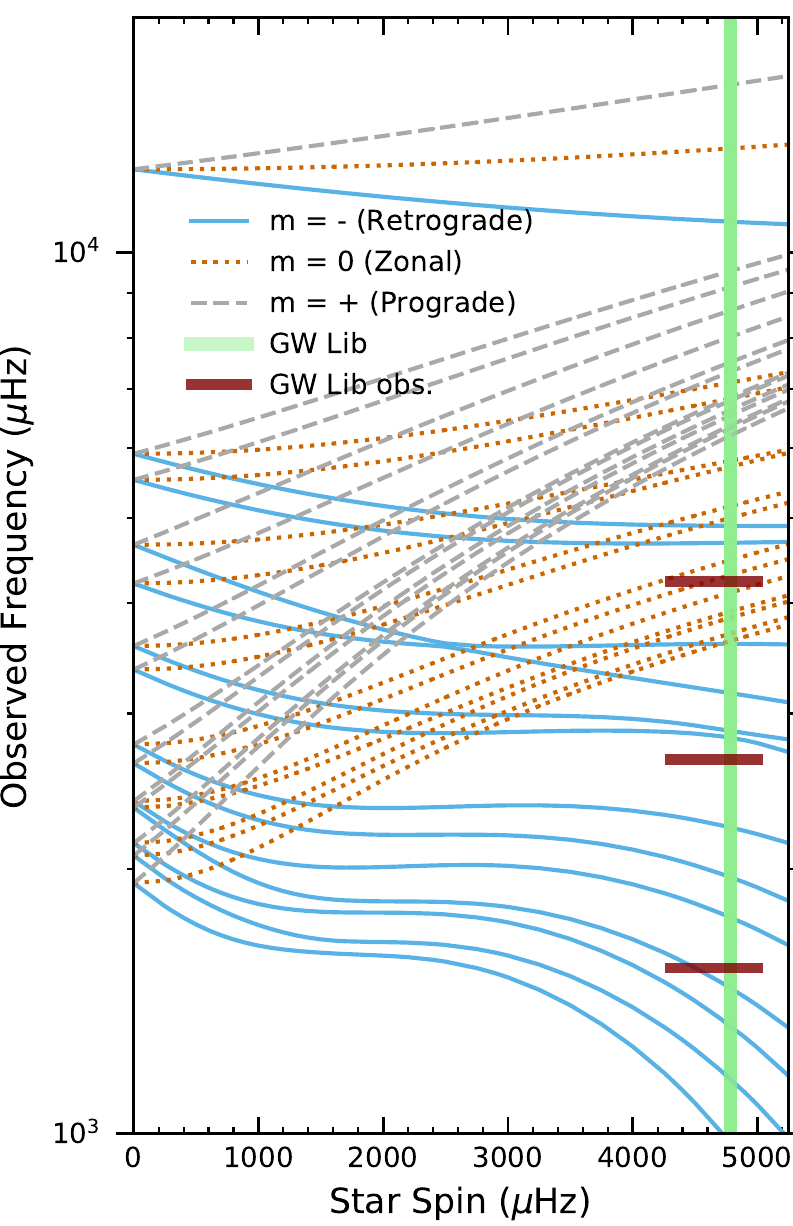}
\caption{Gravity mode frequencies as a function of star’s spin  frequencies  calculated  in  observer's frame (non-rotating frame) of the first 13 modes for the 0.78~M$_\odot$ mass model. Modes shown here are calculated three months after the dwarf nova outburst. Top to bottom indicates lower to higher order modes. The frequencies observed in GW Lib are indicated by the solid maroon horizontal lines on the right panel. From top to bottom: 4237 $\mu$Hz, 2660 $\mu$Hz, and 1543 $\mu$Hz. \label{fig:modes vs spin frequency for 0.78M after three months}}
\end{figure}
Figure \ref{fig:modes vs spin frequency for 0.93M after three months} shows the first 18 $g$-mode frequencies against the star's rotation frequency for the 0.93~M$_{\odot}$ case, computed by Gyre. The vertical solid green band indicates the measured rotation frequency of GW Lib, i.e., 209 seconds. The solid blue lines ($m=-1$) are the retrograde modes, which propagate in the opposite from the direction of the star's spin. The dashed grey lines are the prograde modes ($m=1$), propagating in the same direction of the star's spin. Dotted red lines are the zonal $g$-mode ($m=0$), which is axisymmetric. In the absence of rotation, the eigenfrequencies are (2$\ell$ +1) times degenerate; however, in the presence of rotation, the degeneracy is lifted for a harmonic degree $\ell$, as shown in figure \ref{fig:modes vs spin frequency for 0.93M after three months}. Mode frequencies in the co-rotating (non-inertial) frame, $\omega$, and non-rotating (inertial) frame, $\omega_i$, are related by the spin frequency, $\Omega$, through $\omega_i = \omega+m\Omega$. A distant observer will measure $\omega_i$, as that is the frequency with which a given side of the star (in a non-rotating frame) undergoes a brightness variation. Note that the large-small alternating mode spacing that led to the mode frequencies appearing to be paired in the non-rotating star is still evident here. Also, some modes change which neighbor they are ``paired'' with as the spin frequency increases. At low spin frequencies, much lower than the mode frequency, the modes break into triplets with a small spacing between the different $m$ value modes. Once the spin frequency is higher than a small fraction of the mode frequency, the shift in mode frequency away from the non-rotating value is large. While at small spin frequencies, modes of similar radial order $n_g$ but different $m$ values are grouped together, at large spin frequencies the opposite is more the case: modes of similar $m$ values but different $n_g$ start to become grouped together. The grouping of modes with similar eigenfrequencies that pair together seems exaggerated due to the presence of a sharp interface in our model. We reserve judgement on the robustness of this phenomena until our future work which will include element diffusion during the accretion phase and therefore have a much more realistic core-envelope boundary. One of the reasons that we show both the co-rotating and inertial frame frequencies in Figure \ref{fig:modes vs spin frequency for 0.93M after three months} is that mode driving is quite sensitive to the co-rotating frequency. As discussed in preliminary work by \cite{Townsley_2016}, the highest frequency modes, and therefore the first modes driven as the star cools down through the instability strip of the white dwarf, will be the retrograde ($m=-1$) modes. However, these will not appear at the same frequencies they are driven to due to the interaction of their propagation and the rotation. The three maroon horizontal lines shown on the right Figures 7 and 8 indicate the three observed GW Lib oscillation frequencies. As mentioned in \citet{Townsley_2016}, driving is likely to have the best opportunity to excite the highest frequency modes in the co-rotating frame, consisting mainly of the lowest-order retrograde modes. It is interesting that, when mapped to the observed frequencies, these are the modes closely correspond to the frequencies observed from GW Lib. Note again that we have not fit the model to the GW Lib mode frequencies here by, for example, choosing the  thickness of the accreted layer that gives the best fit.

Modes with different $m$ and $n_g$ values can cross frequencies because their eigenfunctions are distinct. However, two modes with the same $m$ value cannot have a frequency crossing, they instead have an avoided mode crossing. Retrograde modes with order $n_g=11$ and $n_g=12$ show an avoided mode crossing (see in right panel) at around 2500$\mu$Hz, where frequencies of the modes are at the point of the closest approach without actually crossing. This has been studied quite extensively for the sub-giants and red giant pulsating stars \citep{Christensen_Houdek_2010, Deheuvels_and_Michel_2010}. In white dwarfs here, the two resonance cavities whose effective coupling can lead to the crossing feature are the light-element-rich envelope and the C-O core. These are separated by a very localized region of strong abundance gradient, creating an effective cavity boundary.

Figure \ref{fig:modes vs spin frequency for 0.78M after three months} is similar to figure \ref{fig:modes vs spin frequency for 0.93M after three months} but for the 0.78~M$_\odot$ model. The characteristics of the mode behavior are similar to the 0.98~M$_\odot$ model. Retrograde modes of radial orders $n_g=4$ and $n_g=5$ show an avoided mode crossing at a spin frequency of around 2800$\mu$Hz. At small spin frequency, prograde modes with low radial orders show slightly larger frequency spacing between two consecutive order modes, when is compared to the 0.93~M$_{\odot}$ mass model. For example, at star's spin frequency of 2500~$\mu$Hz, the prograde modes $n_g=2$ \& $n_g=3$ show frequency spacing of an approximately 118~$\mu$Hz for the 0.93~M$_{\odot}$ mass model and 404~$\mu$Hz for the 0.78~M$_\odot$ mass model. Similarly at the same spin frequency, the prograde mode orders $n_g=6$ \& $n_g=7$ have frequency spacing of approximately 81~$\mu$Hz and 200~$\mu$Hz, respectively for the larger and smaller mass models. The frequency spacing between the modes may be closely related to the thickness of the outermost accreted layer of the WDs and as well as the mass of the WDs.
\begin{figure}[hb!]
  \includegraphics[scale=1.05]{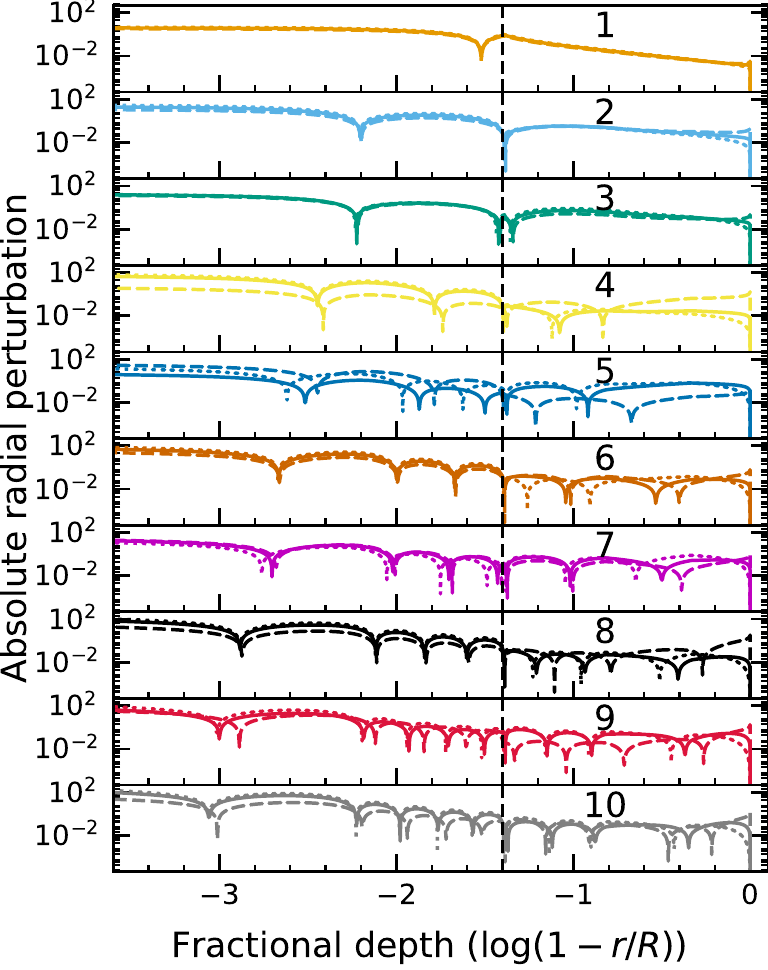}
  \caption{First 10 eigenfunctions for the dipole mode in three different directions: $m=1$ (dashed lines), $m=-1$ (dotted lines), and $m=0$ (solid lines). Vertical axis is the absolute radial displacement perturbations and the horizontal axis is the fractional radius depth from the surface. These eigenfunctions are calculated for the observed GW Lib rotation period (209 seconds) three months after the accretion event for a 0.93~M$_{\odot}$ WD model. \label{fig: eigenfunction_three_months_after_rotation_0.93M} }
\end{figure}

\begin{figure}[ht!]
\includegraphics[scale = 1.]{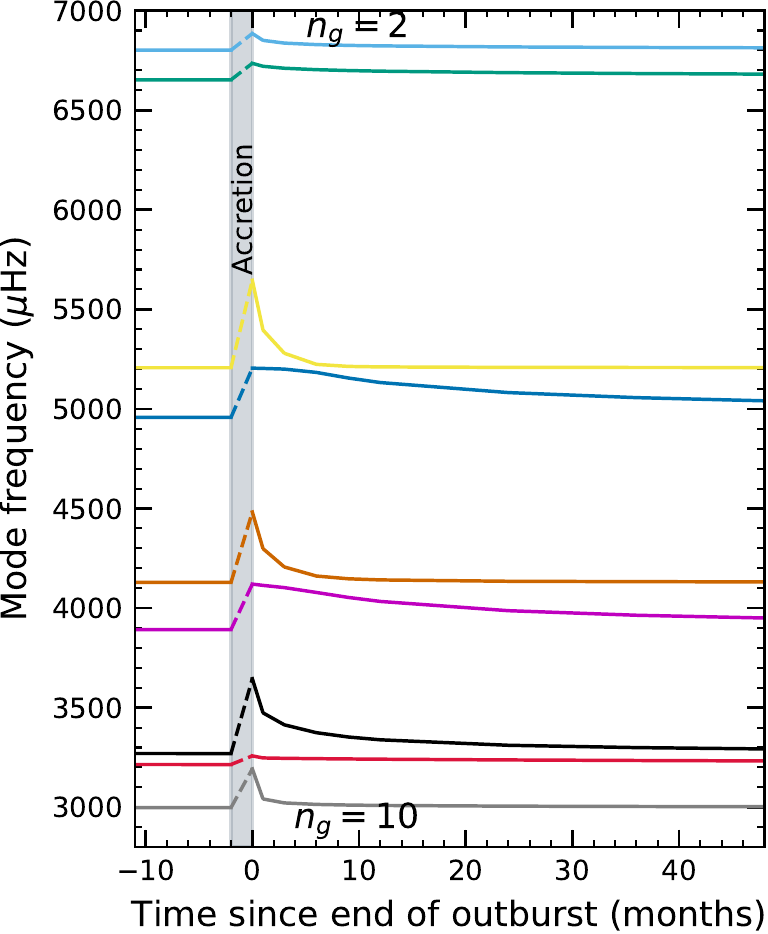}
\caption{$g$-mode frequencies against time since the end of accretion outburst for non-rotating 0.93~M$_\odot$ star. Shown here are the frequencies for the mode orders $n_g=2$ to $n_g=10$. Shaded grey region indicates the two months of accretion time, and dashed lines are to guide the eye and do not represent actual evolution.\label{fig: dwarf nova modes relaxation of GWlib for non rotating case}}
\end{figure}
Figure \ref{fig: eigenfunction_three_months_after_rotation_0.93M} shows the first 10 mode eigenfunctions for 0.93~M$_{\odot}$ model for the rotating white dwarfs three months after the accretion outburst event for the dipole modes ($\ell=1$) for the spin period of 209 s. Eigenfunctions propagating in three directions are indicated with $m=1$ (dashed), $m=-1$ (dotted), and $m=0$ (solid). The vertically dashed perpendicular line indicates the boundary of the core and envelope. Figure \ref{fig: eigenfunction_three_months_after_rotation_0.93M} shows similar eigenfunction behaviour as for eigenfunctions in the non-rotating case. Higher order retrograde modes have a shift in their nodes towards the surface as compared to prograde modes. 

\subsection{Cooling after dwarf nova outbursts} \label{ssec:cooling afer dwarf nova}
\begin{figure*}[ht!]
\includegraphics[scale = 0.98]{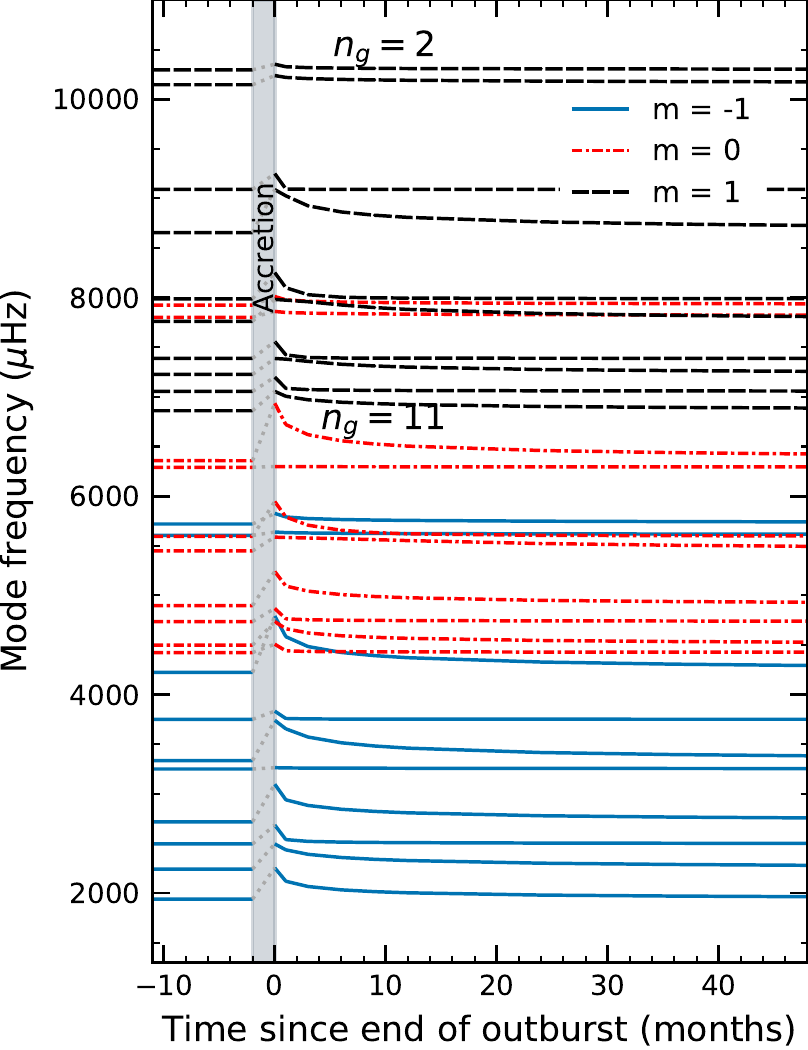}
\includegraphics[scale = 0.98]{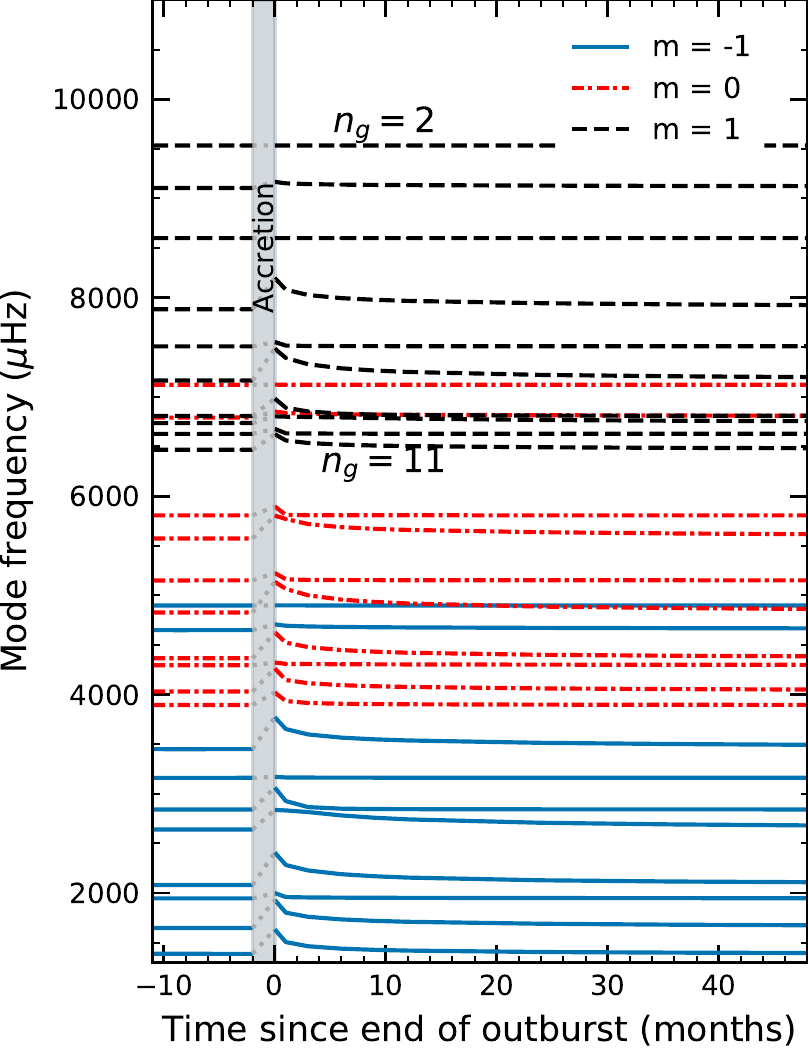}
\caption{Gravity mode frequencies in the inertial frame against time (in months) since the accretion outburst. Left panel is for 0.93~M$_\odot$ and right panel is for 0.78~$M_\odot$ mass models, both for a rotation period of 209 seconds. Top to bottom indicates $n_g$ = 2 to $n_g$ = 11. Three different colors and line styles indicate the three values of the azimuthal eigenfunction index $m$ that make up the dipole triplets.} \label{fig: dwarf nova cooling for GWlib with rotation}
\end{figure*}
Dwarf nova accretion outbursts last for an extremely short timescale as compared to the age of the star. Our computations show that these periods of accretion, at many times the average rate, lead to changes in the propagation of the mode inside the star, resulting in shifts in mode frequencies that are large enough that they should be observable. This is due to the compression, and therefore elevated temperature, of the outer layers created by the sudden addition of new material \citep{Piro_2005}. Observations \citep{Szkody_2012} have shown that the prototype system GW Lib had two large amplitude outbursts separated by about 25 years. 

We begin by showing the time dependence of the normal mode frequencies under the non-rotating assumption. Figure \ref{fig: dwarf nova modes relaxation of GWlib for non rotating case} shows the variations of mode frequencies for $n_g=2$ to $n_g=10$ (top to bottom) gravity modes in time since the end of the accretion outburst for the non-rotating 0.93~M$_{\odot}$ model. The shaded grey region outlines the two months of the dwarf nova accretion event.  
As noted in section \ref{ssec:non-rotating results}, we see mode pairing (see $n_g=2$ \& 3; $n_g=4$ \& 5). However, the variation in frequency with time is very different; the higher frequency mode relaxes faster than the lower frequency one.
For each mode, after the accretion event, there is a few percent increment in frequency, after which the frequency relaxes to its pre-outburst value in a few months time. Lower order modes have a smaller increase after the burst. The $n_g=3$ mode shows a 0.86\% increment from six months before to three months after the accretion outburst, however the $n_g=7$ mode has $\sim$ 5\% increment for the same time frame. Two years after the burst, $n_g=3$ and $n_g=7$ have fallen to a 0.67\% and 2.4\% increment in their mode frequencies respectively.

Figure \ref{fig: dwarf nova cooling for GWlib with rotation} shows the variation in time, after the accretion outburst event, of the frequencies of the $n_g=2$ to $n_g=11$ order modes for 0.93~M$_{\odot}$(left panel) and 0.78~M$_{\odot}$(right panel), for a spin period of 209 seconds (the vertical green line in Figures \ref{fig:modes vs spin frequency for 0.93M after three months} and \ref{fig:modes vs spin frequency for 0.78M after three months}). These are evaluated in the inertial (non-rotating) frame, and so correspond to the frequencies observed through brightness variations by a distant observer. See section \ref{ssec: rotating wd seismolgy method} for relations between quantities in the co-rotating and non-rotating frames. Only modes that correspond to the dipole-like modes ($\ell=1$) in the absence of rotation are shown (see figures \ref{fig:modes vs spin frequency for 0.93M after three months} and \ref{fig:modes vs spin frequency for 0.78M after three months}). The blue solid lines are the prograde modes ($m=1$), the dotted dashed red lines are the retrograde modes ($m=-1$), and the dashed black lines are the zonal modes. Both panels show similar behavior to the non-rotating case, each mode frequency relaxes to its pre-outburst value in approximately 30-40 months. The grey shaded region indicates the two months of strong accretion during the dwarf nova outburst. The minus values on the time axis are taken as before the dwarf accretion event.

The lowest order modes ($n_g=2$ and 3) show a smaller shift in their mode frequencies than higher modes, since low-order modes reside deeper in the star, and thus are less vulnerable to the transient heating and cooling that effects the outer layers of the star. This can be seen by comparing the propagation regions shown in Figure \ref{fig: propagation_diagram_6mb_norot_0.93M} (or the eigenfunctions shown in Figures \ref{fig: eigenfunction_6months_before_and_three_months_after_norotation_0.93M} and \ref{fig: eigenfunction_three_months_after_rotation_0.93M}, see also Figure \ref{fig: Full propagation diagram of 0.93M three months after the burst} below) with the variation in temperature and Brunt-V\"ais\"al\"a shown in Figure \ref{fig:wkb_integrand_brunt_abundance_after_cooling_0.93M}. The heated region where the Brunt frequency is modified by the outburst extends to a pressure of approximately $10^{15}$~erg~cm$^{-3}$, which corresponds to a fractional radial depth of about $10^{-2.2}$, just outside the outer turning point of the $n_g=1$ mode and in a similar location to the outermost node of the $n_g=2$ and 3 modes.

The changes in mode frequency are different both for modes of different radial orders and based on whether a mode propagates with or against rotation (prograde or retrograde). Table \ref{table:table_mode_with_time_change} outlines the frequency change of a few chosen modes from Figure \ref{fig: dwarf nova cooling for GWlib with rotation} at select times after the accretion event. Data shown in the table outline the percentage rise in mode frequency at 0, 3, and 12 months after the end of the accretion outburst. For 0.93~$M_\odot$, the retrograde modes with higher order $n_g$ have a larger rise in their mode frequencies as compared to prograde modes. For example, the $n_g=6$ retrograde mode shows a 12\% rise in frequency just after the accretion outburst event, while the prograde mode shows a 3.30\% rise just after the accretion burst for the same radial order. As mentioned above, the lowest order modes have a smaller rise. For example, the $n_g=1$ modes have have 0.62\% and 0.91\% rise for their retrograde and prograde directions respectively. Moving on to the 0.78~M$_\odot$ model, we see that the higher order modes show an even more significant rise in their mode frequencies. For example, the retrograde $n_g=10$ mode shows a a 17\% rise in frequency just after the accretion event, while the prograde mode of the same radial order shows a 0.76\% rise. 

The timescale on which most mode frequencies relax to their pre-outburst values is small compared to the cooling times between dwarf nova accretion events, 30 years in our example here. Both the offset from the pre-outburst frequency and the rate at which the frequency relaxes are unique to each mode, and these show distinctive patterns across the modes. This presents a hopeful option for identification of the radial order of each mode, which is necessary for robust inference of physical parameters such as accreted layer mass or interior (C/O) composition. In slowly rotating white dwarfs, common practice is to infer radial mode order by identifying a sequence of consecutive or nearly-consecutive radial order modes. The size and variation of the frequency spacing can then be matched to models within reasonable parameter ranges. Due to the large change in the mode frequencies introduced by rotation, sequences of modes of increasing radial order but different $m$ values overlap. This makes the identification of modes through mode sequences more challenging. Distinctive variation during cooling after a dwarf nova accretion event may present a viable alternative, and observing the time-dependent relaxation of modes may provide even more information about the WD interior than the mode frequencies alone.
\begin{table}
\centering
\caption{Mode relaxation with time}
\begin{tabular}{c|c|c|ccc}
\hline \hline
Mass & Radial & Azimuth & Just after & 3 M.a.\footnote{Months after} & 12 M.a.\\
& order & order  & the burst & the burst & the burst\\
($M_\odot$) & ($n_g$) & (m) & (\% rise)\footnote{with respect to pre-outburst value} & (\% rise) & (\% rise)\\ \hline \hline
& 3 & -1 & 0.62 & 0.44 & 0.34\\
& 3 & +1 & 0.91 & 0.612 & 0.445 \\ \cline{2-6}
0.93 &  6 & -1 & 12.17 & 7.12 & 3.77\\
&  6 & +1 & 3.30 & 0.52 & 0.09\\ \cline{2-6}
& 10 & -1 & 11.30 & 6.62 & 3.81\\
& 10 & +1 & 2.05 & 0.26 & 0.13\\ \hline
& 3 & -1 & 1.24 & 0.79 & 0.58\\
& 3 & +1 & 0.67 & 0.43 & 0.31\\ \cline{2-6}
0.78 &  6 & -1 & 7.78 & 0.86 & 0.12\\
& 6 & +1 & 0.58 &0.06 & 0.023\\ \cline{2-6}
& 10 & -1 & 17.13 & 6.82 & 3.77\\
& 10 & +1 & 0.76 & 0.05 & 0.02\\ \hline
\end{tabular}
\label{table:table_mode_with_time_change}
\end{table}

\subsection{Period and Propagation Diagram with Rotation} \label{ssec:periodsandpropagationdiagramswithrotation}
\begin{figure}[ht!]
   \includegraphics[scale=1.]{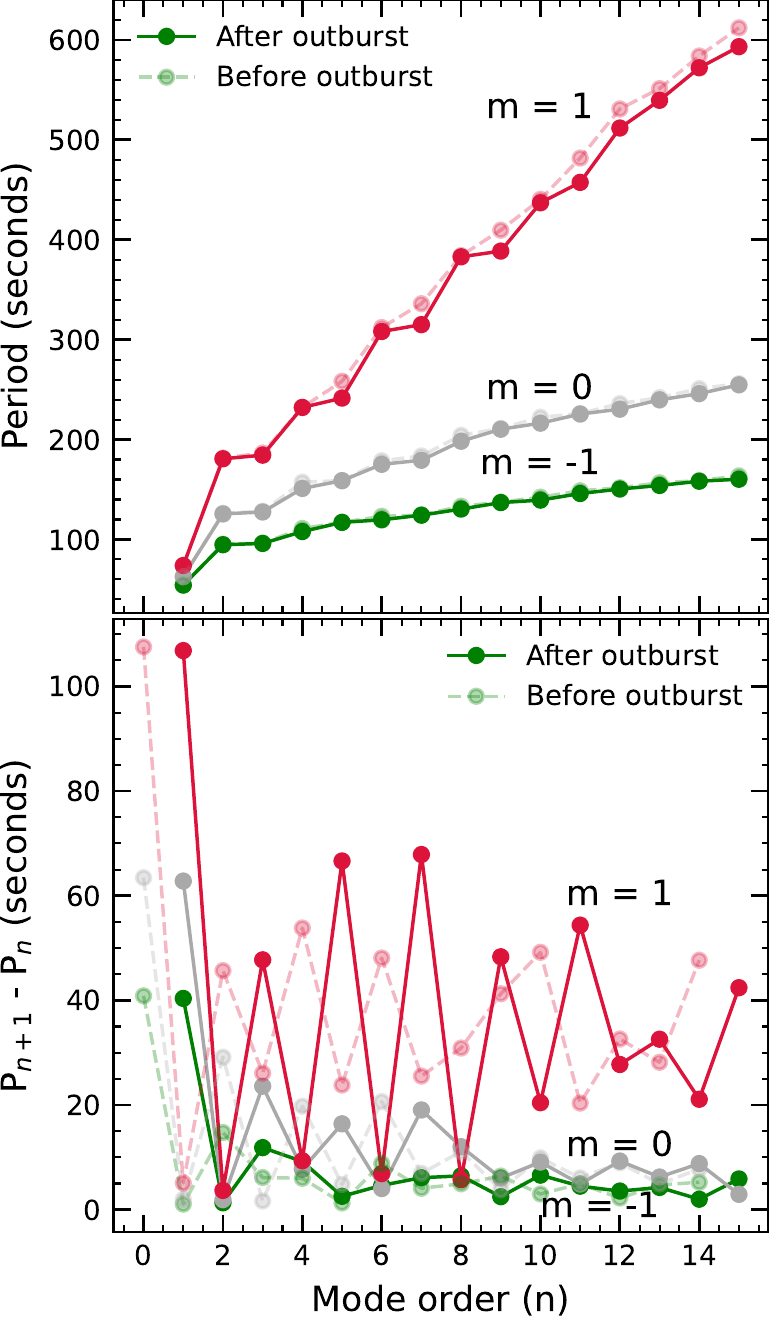}
   \caption{Period (top panel) and period spacing (bottom panel) in the co-rotating frame against mode order for the first 15 modes for a spin period of 209 seconds and mass of 0.93~M$_\odot$ (top panel). Periods are shown six months before (dashed, light) and three months after (solid, dark) the dwarf nova accretion event. Colors indicated $m$ values as labelled.}
   \label{fig: Period against mode order for 0.93M model}
\end{figure}

Figure \ref{fig: Period against mode order for 0.93M model} shows the period (top panel) and period spacing (bottom panel) in the co-rotating frame against the mode order for the first 15 modes. Modes are indicated are with solid circles, with lighter circles and dashed lines indicating values six months before the accretion outburst and darker circles and solid lines indicating three months after the outburst. Prograde ($m=+1$) modes have larger period spacings than the retrograde modes. The average period spacing is related to the total mass, core temperature, and core composition. Although the average period spacing between consecutive mode orders remains similar before and after the outburst, individual prograde modes show shifts in mode period about 50-60 seconds. Retrograde modes show shifts of roughly 10-20 seconds.  
Since Figure \ref{fig: Period against mode order for 0.93M model} shows periods and period spacings in the co-rotating frame, it is necessary to identify the $m$ value of individual modes before observed mode frequencies could be compared to this since the co-rotating frequency is given by $\omega = \omega_i - m\Omega$. This requirement decreases the utility of simple period spacing compared to slowly rotating WDs. The angular eigenfunctions, which are solutions to Laplace's tidal equation, are distinctive, with some being squeezed toward the equator (See e.g. Figure 1 in \cite{Bildsten_Ushomirsky_Cutler_1996}). This may provide some opportunities to distinguish $m$ values based on surface coverage as indicated by time-resolved multi color photometry or, better, spectroscopy. 
\begin{figure}[ht!]
\includegraphics[scale =1.]{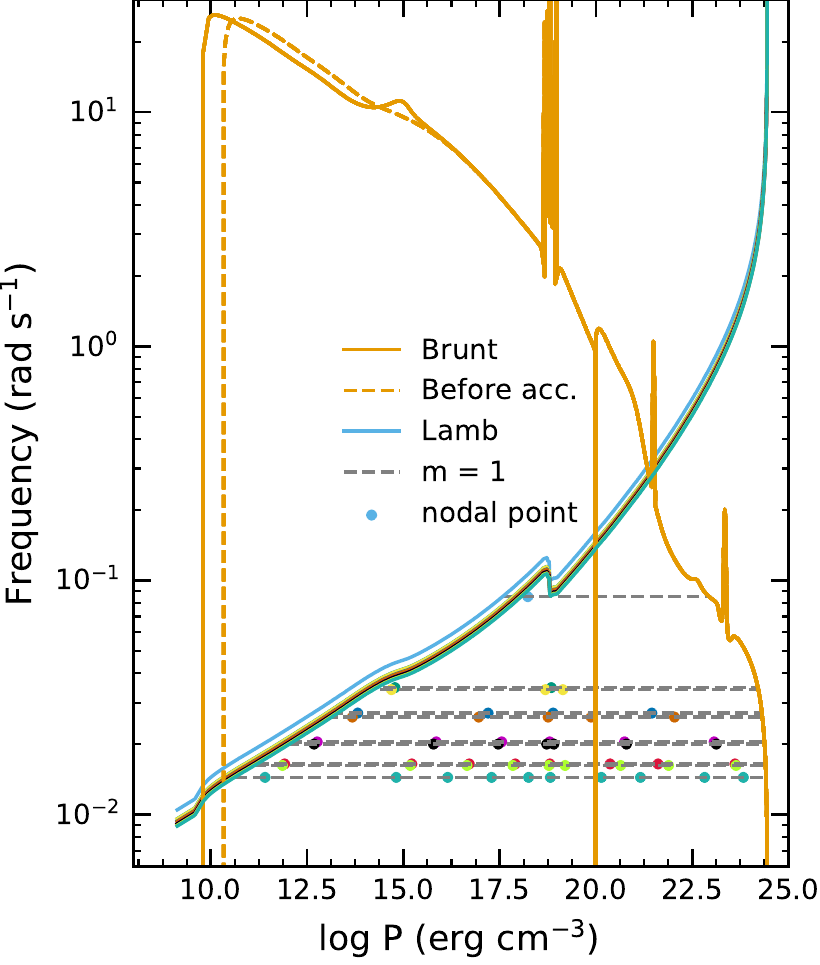}
\caption{Propagation diagram for the 0.93~M$_\odot$ model for the rotating case three months after the accretion event. The solid and dashed orange lines are the Brunt-V{\"a}is{\"a}l{\"a} frequency three months after and just before the accretion event, while the solid lines of various colors indicate the Lamb frequency for individual modes (see text). The frequencies of the ten lowest order $g$-mode are indicated by dashed horizontal lines from top to bottom. Pressure location of the zero crossings of the radial displacement $\xi_r$ are indicated with solid dots of varying colors. \label{fig: Full propagation diagram of 0.93M three months after the burst}}
\end{figure}

Figure \ref{fig: Full propagation diagram of 0.93M three months after the burst} illustrates the full propagation diagram for the 0.93~M$_\odot$ model three months after the accretion event for the first 10 prograde modes. This is very similar to the non-rotating model shown in Figure \ref{fig: propagation_diagram_6mb_norot_0.93M} but with rotation and after the accretion instead of before. The sharp vertical line on Brunt frequency to the left is the base of the convection zone where Brunt frequency goes to zero. The outward boundary of the $g$-mode propagation region is set by the location where the Lamb frequency is equal to the mode frequency (in the co-rotating frame). While in the non-rotating case of the Lamb frequency is proportional to $\ell(\ell+1)$, the equivalent thing in the rotating case is the eigenvalue of Laplace's tidal equation, $\lambda$. (See section \ref{ssec: rotating wd seismolgy method}) Since $\lambda$ depends on the ratio of the mode frequency to the spin frequency, $\omega/\Omega$, the appropriate Lamb frequency lines are different for each mode. Figure \ref{fig: Full propagation diagram of 0.93M three months after the burst} shows ten different Lamb frequency lines for ten different radial order modes. The leftmost is for $n_g=1$ and the rightmost for $n_g=10$. We have chosen to show only prograde modes for clarity. We see that the outer edge of the propagation region, where the Lamb frequency line and the mode frequency line (dashed) intersect, tends to be further inward for higher order modes than one might expect based on the Lamb frequency curve in the non-rotating case. The solid circles denote the zero crossings of the radial displacement eigenfunction $\xi_r$. By comparing to Figure \ref{fig: propagation_diagram_6mb_norot_0.93M} (non-rotating), we see that the outer node location for the prograde modes is fairly close to the outer edge of the propagation region, and thus the movement of the location of the outer turning point can have an appreciable impact on the eigenfrequency of the mode. This also confirms that our mode calculation is behaving correctly, since all zero crossings (nodes) in the eigenfunction are within the propagation region indicated by the $\lambda$-based Lamb frequency. 

\section{Discussion and Conclusion} \label{sec:discussion}

In this work we have laid out how the fundamental $g$-mode structure of accreting white dwarfs should appear.
This includes two features unique to these objects:
(1) the high spin frequencies, comparable to or exceeding the mode frequencies, and
(2) the variation in mode frequencies due to the heating and cooling of the outer layer by bursty accretion (dwarf novae).
While outlining the basic features of seismology for accreting WDs, there are still many challenges to overcome on both the observational and theoretical side.
Here we have focused on a forward-modeling approach rather than trying to fit the observed modes in a particular object.
That is, we selected representative values of the essential parameters of the WD, then carried out an evaluation of the expected mode frequencies and their variation after accretion outbursts.

\subsection{WD seismology under rapid rotation}\label{ssec:wd seismology-rapid rotation}

The 209 s rotation period used here is that observed for GW~Lib, representing a significant spin-up by accretion.
As the shortest $g$-mode periods are around 100~s, and others are longer period, all modes have their frequencies strongly shifted by rotation.
We find for a 209~s spin period that the shift due to rotation is much larger than the separation in period between consecutive radial order modes.
For example, the $n_g=4$ mode in the $0.78$~M$_\odot$ star was shifted from period of about 209~s without rotation to about 120~s, for prograde propagation ($m=+1$).
The average period separation is around 50~s.
The resulting overlapping of mode sequences means that, unlike slowly-rotating WDs, average period separation and deviations from such for observed modes cannot be used directly for seismology.
It is our hope that, with more continuous coverage after accretion outbursts, the variation of individual mode frequencies might be used to identify the azimuthal and radial orders of observed modes and allow a confident seismological fit.

\subsection{Relaxation of mode frequencies after accretion}

This work is the first time variation after accretion outbursts due to surface layer heating and cooling has been computed.
Our representative values were chosen to address the available observations, with the most observed object being GW Lib. We chose two masses, 0.93~M$_\odot$ and 0.78~M$_\odot$, a fairly low accretion rate that gives appropriate surface temperatures, and a spin period (209~s) and time between accretion outbursts (30~yr) similar to GW Lib. 
We have focused on the dipole-like modes, since they should be the most observable, and low radial order, since those modes will have the highest frequencies.

We find that the increase in temperature of the outer layers resulting from the material added during an accretion event, and the subsequent cooling, will shift mode frequencies by up to a few percent (give an example of mode relaxation), with the shift and the rate of return being unique to each mode. This indicates a novel method of seismology that could be applied to these objects with detailed monitoring of the changes in mode frequencies during the months and years following an accretion event (dwarf nova outburst). Rather than $n_g$ being inferred from sets of consecutive order modes, the deviation and relaxation characterstics might be used instead.

\subsection{Current limitations and future work}

While this work presents the expected form of $g$-mode oscillations in accreting WDs, there are several remaining limitations that should be addressed as we move to compare results to observed stars.
The major limitations that we wish to identify and discuss are the need for element diffusion, the simplicity of the evolutionary scenario used here, the, as yet unsolved, role of convective mixing during hydrogen flashes, and the appropriateness of the approximations that allow the traditional approximation for how rotation modifies the modes to be used. We will discuss each of these briefly here.

Element diffusion has been included during the cooling phase of the WD evolution, before the onset of the accretion. This is important to smooth out the WD interior elemental profiles and, therefore, the profile of the buoyancy frequency. However, we have not included element diffusion during the accretion phase, leading to an unrealistic, sharp boundary between the accreted material and the core.  This feature will have a significant impact on the mode frequencies computed for the star. Thus, including element diffusion will be a part of our follow-up work. Although computed mode frequencies may differ, we expect the frequency variations to be similar to those shown in this work.

One of the limitations in this work is that the core temperatures used here are higher than the equilibrium core temperatures found in \cite{Townsley_Bildsten_2004}. As a result, the accretion rate used in this work is lower than that expected for CVs \citep{Shara_2018, Pala_2022}. In our future work, we will improve our treatment of CV evolution, including more realistic core temperatures and accretion rates. One particular challenge is the role of convective mixing during the hydrogen flash. \cite{Jose_2020} show 3-D simulation results and discusses the effect of convective mixing between the hydrogen rich layers and the carbon rich interior on the course of the outburst event. Integrating such results into 1-D stellar evolution models is an area of ongoing research \citep[e.g.][]{Wong_2021}. We refer the reader to \cite{Chiosi_2007} for a review of convective mixing in stars, including theory and observations. 

Here we have made the traditional approximation of rotation (\ref{ssec: rotating wd seismolgy method}), allowing us to obtain a set of $g$-modes for the star. This approximation may not be good in some parts of the star, possibly leading to some differences between the modes that we have found and the actual modes of the star. Therefore, it is important to  treat the normal mode problem without this approximation. In this work, we have not included the importance of $r$-modes. This is another family of modes, in addition to $g$- and $p$- modes (pressure modes), that exist only for non-zero rotation. Therefore, moving beyond the traditional approximation will allow us to treat both $g$- and $r$-modes \citep{Papaloizou_1978, Mukadam_2013, Saio_2019}, as well as modes of mixed character.

\subsection{Conclusions} \label{sec:conclusions}
We have computed $g$-mode frequencies for accreting WDs in the limit of rapid rotation. For each of the calculated modes, there is a few percent rise in its mode frequency after an accretion event, which then relaxes at a unique rate to its pre-outburst value. There are limitations in our treatment of the CV history and the approximation used to compute the $g$-mode frequencies. These will be addressed in our follow-up work.

The relevant files and datasets are available online at \url{https://doi.org/10.5281/zenodo.7884172}.
\acknowledgments
We thank Ken Shen, Alan Calder and Sam Boos for the useful discussions. We also would like to thank Spencer Caldwell and Broxten Miles for their earlier work on setting up the nova simulation. This work was supported under programs HST-GO-15072, HST-GO-16069, and HST-AR-16638, through the Space Telescope Science Institute, which is operated by the Association of Universities for Research in Astronomy, Incorporated, under NASA contract NAS5-26555. Support for these programs was provided through a grant from the STScI under NASA contract NAS5-26555.

Software: \texttt{MESA} \citep[][\url{mesa.sourceforge.net}]{Paxton_2011, Paxton_2013, Paxton_2015, Paxton_2018}, 
\texttt{GYRE} \citep[][\url{https://gyre.readthedocs.io/en/stable/}]{Townsend_2013, Townsend_2018}


\bibliography{references}{}
\bibliographystyle{aasjournal}

\end{document}